\documentclass[%
 aip,
 amsmath,amssymb,
 reprint,%
]{revtex4-1}

\usepackage{graphicx}% Include figure files
\usepackage{dcolumn}% Align table columns on decimal point
\usepackage{bm}% bold math
\usepackage[utf8]{inputenc}
\usepackage[T1]{fontenc}
\usepackage{mathptmx}
\usepackage{etoolbox}
\usepackage{sidecap}
\usepackage{epstopdf}
\usepackage{epsfig}
\usepackage{amsmath}			
\usepackage{amsfonts}			
\usepackage{amssymb}			
\usepackage{latexsym}			
\usepackage{color}
\usepackage{soul}
\usepackage{ulem}
\usepackage{multirow}

\makeatletter
\def\@email#1#2{%
 \endgroup
 \patchcmd{\titleblock@produce}
  {\frontmatter@RRAPformat}
  {\frontmatter@RRAPformat{\produce@RRAP{*#1\href{mailto:#2}{#2}}}\frontmatter@RRAPformat}
  {}{}
}%
\makeatother
\begin{document}

\preprint{AIP/123-QED}

\title{Accounting for speckle scale  beam-bending in classical ray tracing schemes for propagating realistic pulses in indirect drive ignition conditions}
\author{C. Ruyer}\email{charles.ruyer@cea.fr}
\affiliation{CEA, DAM, DIF, F-91297 Arpajon, France}
\affiliation{Universit\'e Paris-Saclay, CEA, LMCE, 91680 Bruy\`eres-le-Ch\^atel, France}
\author{P. Loiseau}
\affiliation{CEA, DAM, DIF, F-91297 Arpajon, France}
\affiliation{Universit\'e Paris-Saclay, CEA, LMCE, 91680 Bruy\`eres-le-Ch\^atel, France}
\author{G. Riazuelo}
\affiliation{CEA, DAM, DIF, F-91297 Arpajon, France}
\affiliation{Universit\'e Paris-Saclay, CEA, LMCE, 91680 Bruy\`eres-le-Ch\^atel, France}
\author{R. Riquier}
\affiliation{CEA, DAM, DIF, F-91297 Arpajon, France}
\author{A. Debayle}
\affiliation{CEA, DAM, DIF, F-91297 Arpajon, France}
\affiliation{Universit\'e Paris-Saclay, CEA, LMCE, 91680 Bruy\`eres-le-Ch\^atel, France}
\author{P. E. Masson-Laborde}
\affiliation{CEA, DAM, DIF, F-91297 Arpajon, France}
\affiliation{Universit\'e Paris-Saclay, CEA, LMCE, 91680 Bruy\`eres-le-Ch\^atel, France}

\date{\today}% It is always \today, today,

\begin{abstract}
We propose a semi-analytical modeling of smoothed laser beams deviation induced by  plasma flows. Based on a Gaussian description of speckles, the model includes spatial, temporal and polarization smoothing techniques, through fits issued from  hydrodynamic simulations with a paraxial description of electromagnetic waves.
This beam bending model is then included in a ray tracing algorithm, and carefully validated.
When applied as a post-process to the propagation of the  inner cone in a full-scale simulation of a NIF experiment, the beam bending along the path of the laser affects the refraction conditions inside the hohlraum and the energy deposition, and could explain the anomalous refraction measurements, the so-called glint observed in some NIF experiments.
\end{abstract}

\maketitle

\section{Introduction}
Many experiments conducted in multi-kilojoule laser facilities, whether they concern astrophysical phenomena, high-energy-density physics \cite{Drake2006} or inertial confinement fusion \cite{Lindl_2004,Cavailler_2005,PRL_Lan_2021},
bring critical insights into  matter under extreme  conditions. 
In these facilities, energetic laser beams are used to heat  and compress the matter to millions of degrees and  Mega to Gigabar pressure. 
The predictability of these experiments requires a precise understanding of the laser plasma interaction such as 
the mechanisms responsible for the energy deposition, or the growth of various instabilities such as stimulated Raman or Brillouin scatterings  \cite{Shen_1965,Forslund_1973,POP_Liu_2009,hao_2013}, cross-beam energy transfer \cite{hao_2016}, two-plasmon decay  \cite{Dubois_1995,Russell_2001}, collective scattering \cite[]{Dubois_1992,POP_Xiao_2019,MRE_Qiu_2021}, or self-focusing  \cite{Wagner_1968}.

Furthermore, optical smoothing techniques combining  Random-Phase-Plates (RPP) and  smoothing by spectral dispersion (SSD), available on the Laser M\'ega Joule (LMJ), the National Ignition Facility (NIF), the laboratory for laser energetics (LLE), or SG-III-class lasers, improve the laser intensity profile \cite{Kato_1984,Skupski_1989}.
These techniques reduce  the spatial (through RPP) and temporal coherence (through SSD) of the  light,  resulting in intensity fluctuations on the scale of a few wavelengths lasting a few picoseconds: the so-called speckles. Polarization smoothing (PS), which consists in splitting the pulse in two uncorrelated superimposed beams with perpendicular polarizations, may also be used. 
Together, these smoothing techniques will in turn affect the laser-plasma interactions on macroscopic scales such as the energy deposition region  \cite[]{POP_Delamater_1996,Huser_2009}, the scattering direction  of the light wave \cite{Epstein_1986,PRL_Moody_96,POP_Debayle_2018,POP_Duluc_2019,Yin_2019,POP_Huller_2020} or the amount of expected reflectivity \cite[]{POP_Laffite_2010,POP_Masson_2016,Glize_2017,Winjum_2019}. Accounting for the hot-spot dynamics thus requires to conciliate the sub-micron and sub-picosecond physics of the RPP/SSD beam with the millimeters size and nanoseconds duration of the experiments. 
In this context, a hydrodynamic description of the plasma can be coupled with an approximated Maxwell solver  \cite{Berger_1995,Still_2006,Loiseau_2006, Huller_2006}. However this formalism fails to capture self-consistently the acoustic waves Landau damping or other kinetic effects and may appear to be numerically unsuitable whenever  multi-dimensional effects,  solid-density physics or radiative phenomenons arise.

In such systems, hydrodynamic codes rest on a rough description of light,  such as the classical ray tracing scheme \cite[]{Hydra,Lared-H,POP_Zhang_2014,Lefebvre_2018} which, in its native form, only captures the wave refraction and basic energy deposition.
The accurate description, by the ray tracing scheme,  of simple  light quantities  such as the intensity  of the wave or its local spectrum, is still an active area of research  \cite{Egorchenkov_2001, Colaitis_2014, Strozzi_2017,POP_Colaitis_2019}.
Although a large effort is made in improving these schemes by including back or side scattering of the light caused by wave mixing processes \cite[]{Strozzi_2017,POP_Debayle_2019},  
modeling the speckle scale physics  remains vastly unexplored \cite[]{PRL_Hinkel_1996,POP_Grech_2006,PRL_Grech_2009,PRL_Rousseaux_2016}.
 
This study  addresses the beam bending  of  a laser beam \cite{POP_Hinkel_1998,POP_Bezzerides_1998,POP_Rose_96,PRL_Montgomery}, provided other instabilities, such as filamentation and forward Brillouin scatterings are negligible.
It occurs when the laser driven density fluctuations are advected by a flow, resulting,  due to a wave-guide effect, to the deflection of the electromagnetic wave off its original propagating axis, toward the flow direction. 
As this effect may take place in a perfectly homogeneous plasma, the beam deflection may add up to the well-known refraction of light caused by density gradients. 
In section II, we will first briefly recall the kinetic and fluid  modeling of Ref \cite[]{POP_Ruyer_2020}, which predict the deflection angle of a Gaussian laser pulse, in both the transient and asymptotic regimes of the plasma density response.
The model is then extended, in section III, to the centroid deviation of a spatially (using RPP) and temporally (with SSD) incoherent laser pulse with the effect of polarization smoothing by fits issued from  three-dimensional (3D) %Hera \cite{HERA_Jourdren_2005,Loiseau_2006} and 
Parax \cite[]{POP_Riazuelo_2000} simulations.  
%The planar geometry simulations will be used to validate our model over many laser bandwidth and plasma parameters. The more expensive 3D simulations will serve to fit  our calculations on the different smoothing techniques relevant to NIF or LMJ-class laser facilities. 
Our model  is included into 
a  ray tracing description of the beam propagation 
and compared successfully to  our numerical results, 
in section IV. 
Then,
the laser propagation in realistic indirect drive ICF plasma conditions (NIF shot N181209 from the hybrid B campaign \cite[]{POP_Hurricane_2019,POP_Kritcher_2020,POP_Zylstra_2020,POP_Hohenberger_2020}), is addressed and evidences a sensible impact of the flow-induced deviation on the light energy path and energy deposition. 
The last section gather our concluding remarks and perspectives. 

The SI unit system is used throughout this publication, the Boltzmann constant is dropped, the temperatures are given in eV and vectors are noted in bold symbols. 

\section{The beam bending of a spatio-temporally incoherent laser pulse}\label{sec:gauss}
\subsection{Beam bending with the transient regime}
In reference \cite{POP_Rose_96}, proof is made that in the small angle limit,  the transverse averaged beam wavevector, $ \langle \mathbf{k}_\perp\rangle_{\mathbf{k}_\perp}$ may be related to the plasma electron density fluctuations, $\delta n_e$, through 
  \begin{align}
\frac{1}{k_0} \frac{d \langle \mathbf{k}_\perp \rangle_{\mathbf{k}_\perp}    }{d x} \cdot \frac{ \mathbf{v}_d }{ \vert \mathbf{v}_d \vert } &
%= \frac{d\theta}{dx} 
=\frac{-1}{2}\frac{n_e}{n_c}
\left\langle\nabla_\perp\frac{\delta n_e}{n_e }\right\rangle_\perp \label{eq:bbrate0} \, ,\\
\left\langle X \right\rangle_\perp &= \frac{ \int d\mathbf{r}_\perp  X(\mathbf{r})   I(\mathbf{r}) }{ \int d\mathbf{r}_\perp    I(\mathbf{r})} \label{eq:ave}\, , \\
\left\langle X \right\rangle_{\mathbf{k}_\perp} &= \frac{ \int d\mathbf{k}_\perp  X(\mathbf{k})   I(\mathbf{k}) }{ \int d\mathbf{k}_\perp   I(\mathbf{k})} \label{eq:avek}\, ,
\end{align}
 where we introduced $k_0=2\pi/\lambda_0$, $n_c$, $\mathbf{v}_d$,  the main laser wavevector, critical density and flow velocity, respectively.  Moreover, $\mathbf{r}_\perp$ and $\mathbf{k}_\perp$ are the position and wavevector transverse to the main laser $x$ direction. The transverse averages in real and Fourier space, $\left\langle \cdot \right\rangle_\perp$ and $\left\langle \cdot \right\rangle_{\mathbf{k}_\perp}$, respectively, are defined through Eqs. \eqref{eq:ave} and \eqref{eq:avek}. 
We assume hereafter a perfectly homogeneous plasma.
The normalized density fluctuations dependence on time and position, $\delta n_e(t,\mathbf{r})/n_e$, will be assumed to follow  the  linearized hydrodynamic equations. In the plasma rest frame and after a transverse Fourier transform [defined hereafter as $ f(\omega,\mathbf{k})=\int f(t,\mathbf{r})e^{-i\mathbf{k}\cdot\mathbf{r}+i\omega t} d\mathbf{r}dt$], the relation between the density fluctuations and the laser intensity $I(\mathbf{k})=I_0\,g(\mathbf{k})$ verifies, 
\begin{align}
    \left(\frac{\partial^2}{\partial t^2} + 2\gamma_0 \vert \mathbf{k}\vert c_s \frac{\partial}{\partial t} +c_s^2 \mathbf{k}^2\right)\frac{\delta n_e}{n_e } = \nonumber\\ 
   A_{\vert \mathbf{k}\vert}   \frac{c_s^2 \mathbf{k}^2I_0}{n_cv_gT_e} \frac{\alpha_k}{\alpha_f}g(\mathbf{k}) e^{i\mathbf{v}_d \cdot \mathbf{k}t}\, . \label{eq:4}
\end{align}
Here, we introduced the sound speed, $c_s=[(Z_iT_e+3T_i)/m_i]^{1/2}$,  the  electron/ion, mass and  temperature, $ m_{e/i}$ and $T_{e/i}$,  the normalized acoustic Landau damping rate, $\gamma_0=\nu/\vert k\vert c_s$, the laser group velocity, $v_g=c\sqrt{1-n_e/n_c}$ and  light speed in vacuum, $c$.
The ratio, $\alpha_k/\alpha_f$, where 
\begin{align}
 \alpha_ \mathrm{k} [M_0\cos(\theta)]   & = \frac{-\mathcal{Z}'( \zeta_e) }{2}\frac{\sum_i \mathcal{Z}'( \zeta_i)\frac{  Z_iT_e}{ T_i } \frac{  Z_in_e}{ n_i }  }{\mathcal{Z}'( \xi_e)+ \sum_i \mathcal{Z}'( \zeta_i)\frac{  Z_iT_e}{ T_i } \frac{  Z_in_e}{ n_i }   } \, ,\label{eq:drakek}
 \\
\zeta_{e/i} &=  \frac{-\mathbf{k}_\perp\cdot\mathbf{v}_{d} }{\vert \mathbf{k}_\perp \vert } \sqrt{ \frac{ m_{e/i} }{ 2T_{e/i} }  }   \label{eq:xi} \, , \\
 \alpha_f[M_0\cos(\theta)]   & =\frac{\kappa}{1-\mathbf{M}_0^2\cos^2\theta +2i\gamma_0 \vert\mathbf{M}_0\vert \cos\theta}  \, ,  \label{eq:drakeh} \\
 \cos(\theta)& = \frac{\mathbf{k}_\perp\cdot\mathbf{v}_{d} }{\vert \mathbf{k}_\perp \vert \vert \mathbf{v}_{d}\vert  } \, , \label{eq:hydro} \\
 \kappa &= \frac{Z_iT_e}{m_ic_s^2} \, ,
\end{align}
introduced in Ref. \cite[]{POP_Ruyer_2020}, is an ad hoc correction that ensures convergence to the  kinetic asymptotic limit for highly Landau damped plasmas. 
We also introduced the Mach number $\mathbf{M}_0= \mathbf{v}_d/c_s$ and $\mathcal{Z}'$, the first order derivative of the plasma dispersion function \cite{Fried_Gell-Mann_1960}.
Finally, the factor $A_k$, on the r.h.s of Eq. \eqref{eq:4},  accounts for nonlocal thermal effects on the density fluctuation amplitude~\cite[]{POP_Brantov_1998,Bychenkov_2000} and writes,
\begin{align}
     A_k(u)   &= \frac{1}{2} +Z_i\left( \frac{0.074}{u^2}+ \frac{0.88}{u^{4/7}} + \frac{2.54}{1+5.5u^2} \right) \Omega\, ,\nonumber \\ 
     u &=\vert \mathbf{k}_\perp \vert\lambda_\mathrm{mfp} \sqrt{Z_i}\label{eq:nl}\, , \nonumber\\
     \Omega &= 1\,  \mathrm{if}\, \vert \mathbf{k}_\perp\cdot \mathbf{v}_d\vert/\nu_{ei}<1 \, \mathrm{and}\, 0 \, \mathrm{elsewhere, } 
\end{align}
where $\lambda_\mathrm{mfp}$ is the electron mean free path and where $A_{\vert \mathbf{k}\vert}=1/2$ in the collisionless limit. 
As mentioned in Ref. \cite[]{Berger_2005}, this correction can be used provided the acoustic wave frequency, $\vert \omega_s\vert= \vert kv_d \vert $, remains smaller than the electron-ion collisionnal frequency, $\nu_{ei}$ (momentum exchange). This condition is contained in the factor $\Omega$ which ensures that $A_k$ recovers its collisionless value when it is not fulfilled. Easily introduced in our theoretical estimates, such factor remains far more challenging to implement  in a self-consistent numerical modeling where other laser-plasma effects are accounted for, in which case $\Omega$ is set to unity (see Sec. \ref{sec:gauss3D}).
Likewise, the Landau damping factor, $\gamma_0$ in Eqs. \eqref{eq:drakeh} and \eqref{eq:ftransient}, may be expressed in the collisionless limit (Landau) [Eq. (10) in Ref. \cite{POP_Ruyer_2020}], as in Sec. \ref{sec:parax}.
When collisionnal effects are not negligible, the fit as introduced in Ref. \cite[]{LPB_casanova_1989} will be used and leads to a dependence of $\gamma_0$ on $\vert \mathbf{k}\vert $, as in Sec. \ref{sec:icf}.

In the laboratory frame, the solution of Eq. \eqref{eq:4} for $\gamma_0<1$, with the initial conditions $\delta n_e(t=0)=\partial_t \delta n_e(t=0) = 0$, reads,  
\begin{align}
    \frac{\delta n_e(\mathbf{k},t)}{n_e} = &\alpha_k A_{\vert \mathbf{k}\vert}\frac{I_0 g(\mathbf{k})  }{n_cv_gT_e} f(\mathbf{k},t)\, , \\
    f(\mathbf{k},t)=&1+a_+e^{g_+c_s\vert\mathbf{k}\vert t}- a_-e^{g_-c_s \vert\mathbf{k} \vert t}\, , \label{eq:ftransient} \\
    a_\pm =&\frac{-iM_0\cos(\theta) -\gamma_0\mp i\sqrt{1-\gamma_0^2}}{2i\sqrt{1-\gamma_0^2}} \, , \\
    g_\pm =& -\gamma_0\pm i\sqrt{1-\gamma_0^2} -iM_0 \cos(\theta) \, .
\end{align}
Here, $M_0=\vert \mathbf{M}_0\vert $ where $\mathbf{M}_0$ and $ \mathbf{v}_d$ are assumed to be normal to the main laser $x$-axis. %$D$ represents the number of dimensions normal to the main laser axis. 
The combination of the above density fluctuations, written in the Fourier space  with Eq. \eqref{eq:bbrate0}   implies an inverse Fourier transform and leads to,
\begin{align}
  \frac{d\theta }{dx}= \frac{1}{k_0} \frac{d \langle \mathbf{k}_\perp \rangle_\perp    }{d x} \cdot \frac{ \mathbf{v}_d }{ \vert \mathbf{v}_d \vert }
=\frac{-1}{2}\frac{n_e}{n_c} \frac{I_0 }{n_cv_gT_e} \times \nonumber \\
 \int \frac{d^2 \mathbf{k}}{(2\pi)^2} \frac{ i \mathbf{k}\cdot\mathbf{v}_d }{\vert \mathbf{v}_d \vert} \alpha_k A_{\vert \mathbf{k}\vert}g(\mathbf{k})  f(\mathbf{k},t)\left\langle  e^{i \mathbf{k}\cdot  \mathbf{r}_\perp}\right\rangle_\perp\, . \label{eq:bb1}
\end{align}
Hence, the estimation of $\left\langle e^{i \mathbf{k}\cdot  \mathbf{r}_\perp}\right\rangle_\perp$ with Eq. \eqref{eq:ave} leads to the final beam bending rate and it depends on the Fourier transform of the  transverse intensity profile, $g(\mathbf{k})$. 
References  \cite[]{POP_Rose_96,POP_Hinkel_1998} assume Gaussian beams while Ref. \cite[]{POP_Rose_97} addresses RPP beams, in the asymptotic regime only.  
In order to avoid any assumption on the SSD beam temporal spectrum, as in Refs. \cite[]{POP_Ghosal_1997,POP_Rose_Ghosal_98}, 
 we will simply  relate the averaged SSD beam centroid deflection to the speckle  contributions using  Gaussian transient regime estimates \cite[]{POP_Ruyer_2020}.

\subsection{ Gaussian speckle in  three dimensions }\label{sec:gauss3D}

\begin{figure}
\begin{tabular}{c}
%(a) $\log_{10}(I \, [\mathrm{W/cm^2}])$\\
%\includegraphics[scale=0.46]{I_Hem1_100ps.pdf}\\
%\includegraphics[scale=0.39]{Figure_1.eps} 
(a) $d\theta / dx$\\
\\\includegraphics[scale=0.45]{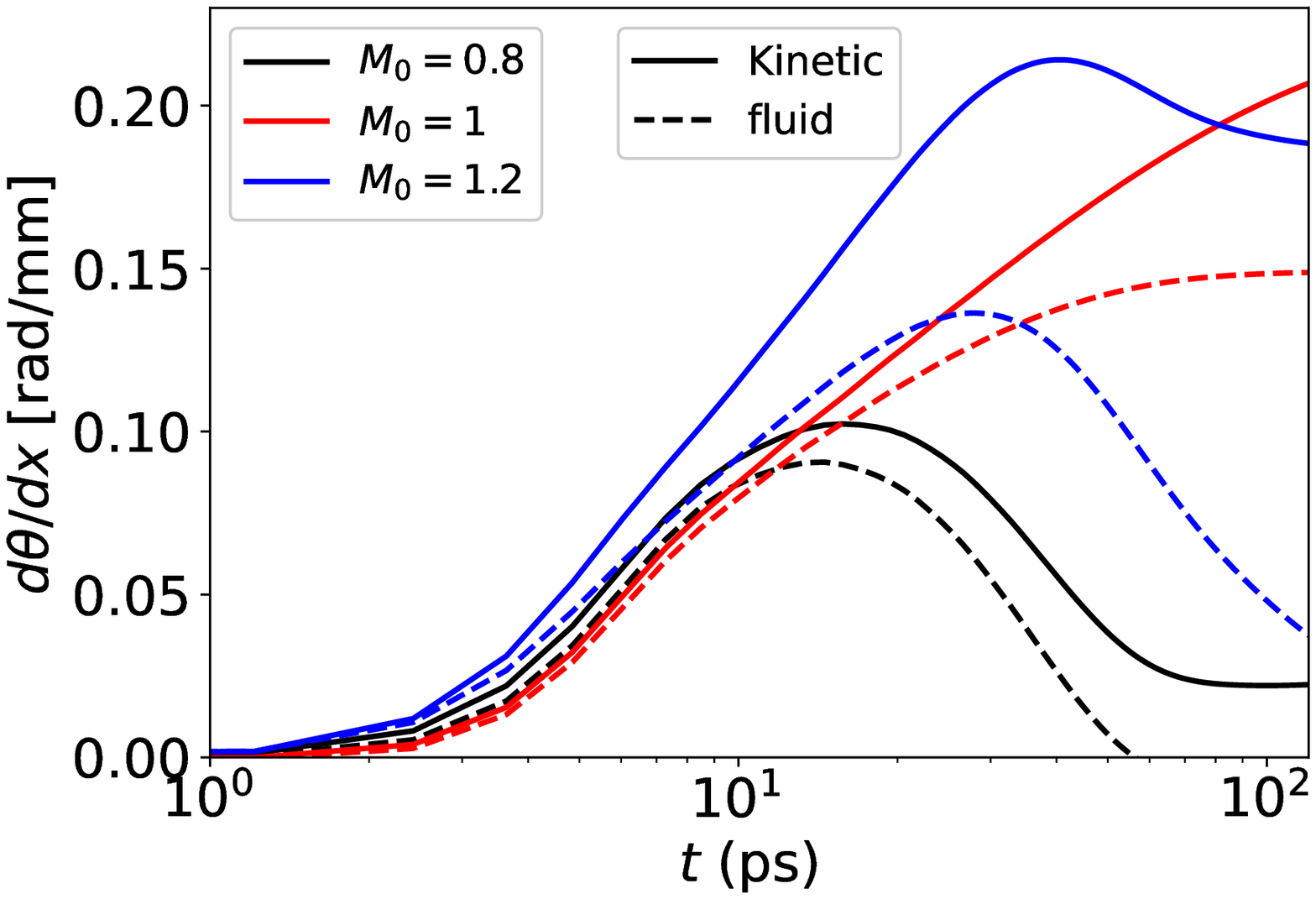} \\
(b) $\langle d\theta/dx\rangle_{\tau_\mathrm{SSD} }$, $\tau_\mathrm{SSD}=7\, \rm ps$  \\
\\\includegraphics[scale=0.45]{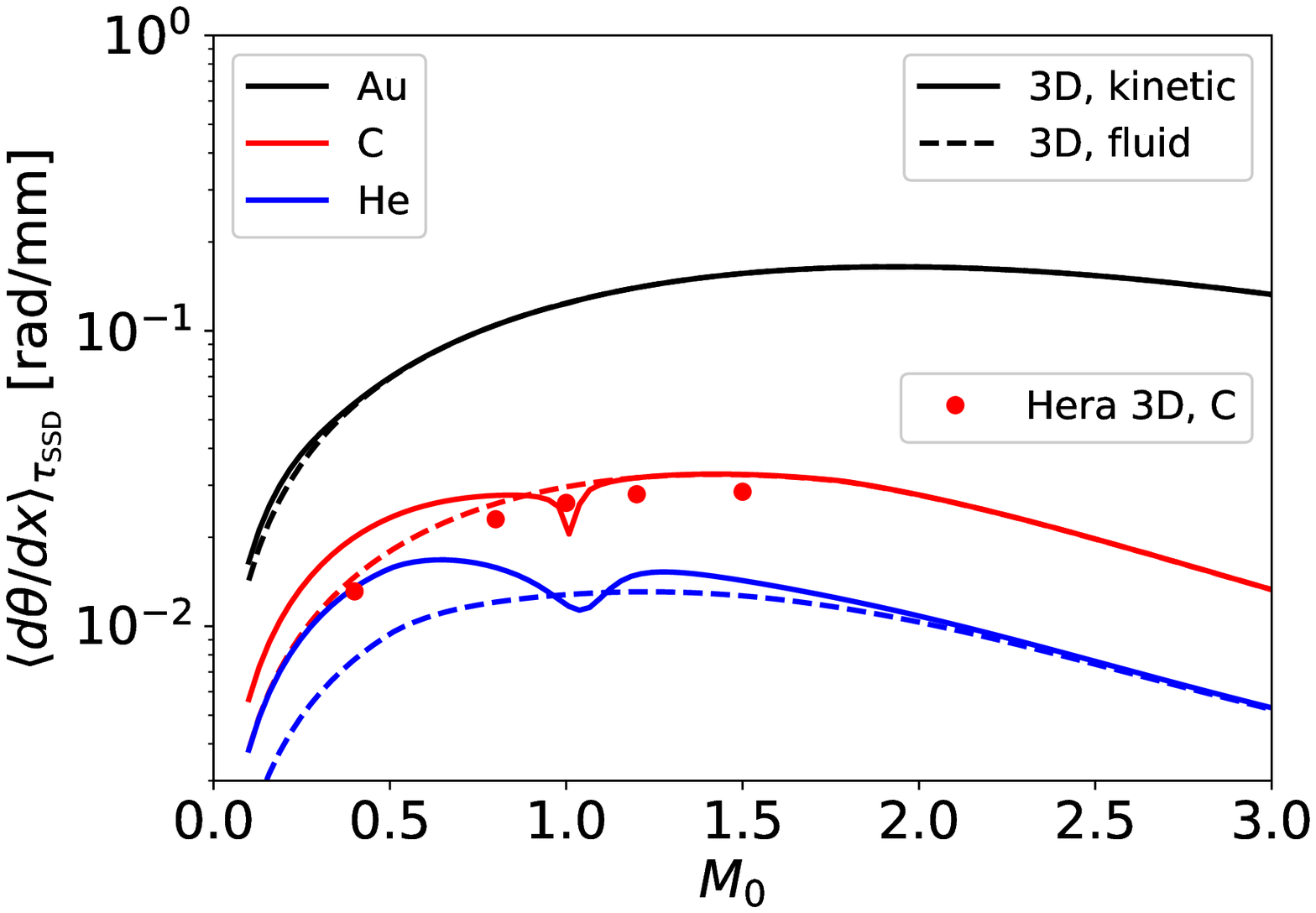} \\
\end{tabular}
\caption{ \label{fig:2d3dnl}
%(a) Intensity profile at $100\,\rm ps$ of a laser beam with $f_\sharp=6.5$, $I_0=4\times 10^{14}\,\rm W/cm^2$, $\lambda_0=0.35\, \rm \mu m$, a SSD with $\omega_m=2\pi14.25\,\rm GHz$ and  $\Delta=3$. The beam propagates through a homogeneous He$^{2+}$ plasma with $T_e=2\,\rm keV$, $T_i=500\,\rm eV$ and  $n_e/n_c=0.1$. The position of the laser centroid is marked by the black plain line.
(a) Deflection rate as a function of time as predicted by Eqs. \eqref{eq:bbf} and \eqref{eq:G3}  for a C$^{6+}$ plasma for various Mach numbers in the fluid [Eq. \eqref{eq:drakeh}] and kinetic formalism [Eq. \eqref{eq:drakek}].
(b) Gaussian deflection rate averaged over $7\,\rm ps$ as predicted by  Eqs. \eqref{eq:bbssd} and  \eqref{eq:G3m} (using $S=1$ and $\beta=1$)  for a He$^{2+}$ (blue lines), C$^{6+}$ (red lines) and Au$^{50+}$ (black lines) plasmas as a function of the Mach number. Three dimensional predictions obtained with Hera are superimposed as red circles (see appendix \ref{sec:hera3d} for details).
(b,c) The laser  has an averaged intensity of $I_0=1\times 10^{15}\,\rm W.cm^{-2}$ and $\lambda_0=0.35\, \rm \mu m$, $f_\sharp=8$,  $T_e=2.5\,\rm keV$, $T_i=1$ keV, $n_e=0.1 n_c$.
}
\end{figure}
Hence,  using $g(\mathbf{k})=\exp(-\mathbf{k}^2\sigma^2/8)$ with $\sigma =\lambda_0f_\sharp$   and assuming a $y$-aligned drift velocity, the deflection rate may be recast as 
  \begin{align}
  \frac{d\theta}{dx}=  -\frac{n_e }{n_c} \frac{ I_0 }{ 2 v_g n_c T_e } \frac{1}{\sigma}\mathcal{G}(t)\, .\label{eq:bbf} 
  \end{align}
  with,  
\begin{align}
  \mathcal{G}=\frac{\sigma^3}{8\pi} \Im \int_{-\infty}^\infty dk  k^2  A_{k}e^{-k^2\sigma^2/4} \times\nonumber\\
\int_0^\pi d\theta \alpha_{k/f}[M_0\cos(\theta)]\cos(\theta) f(k,t) \, .\label{eq:G3}
\end{align}

Figure \ref{fig:2d3dnl}(a) illustrates the temporal evolution of a single three dimensional Gaussian speckle  of focal, $f_\sharp=8$, and wavelength, $\lambda_0=0.35\,\rm\mu m$, in a fully ionized carbon plasma as predicted by Eqs. \eqref{eq:bbf} and \eqref{eq:G3}. Although the deflection is larger for a flow with $M_0=1$ (red lines) in the asymptotic limit ($t> 100 \,\rm ps$), the bending at resonance remains slightly weaker than the case $M_0=0.8$ (as black lines) or  $1.2$ (as blue lines) during a part of the transient regime ($<10 \,\rm ps$). As a corollary result, the resonance of the hot spot bending rate around a Mach number of the order of unity is not noticeable during the transient. 

\subsection{ Spatially and temporally smoothed laser beam } \label{sec:gaussssd}
The  propagation of a spatially and temporally smoothed laser beam in a flowing plasma, as shown in Ref. \cite[]{PRL_Hinkel_1996}, 
presents a deflection of its centroid that can be related to the hot spots contribution.
Assuming the speckles are independent and of Gaussian form, we may now relate the time-averaged beam centroid deflection rate to the  contribution of the speckles ($s$-subscript) through, 
\begin{align} 
  \frac{d\theta_\mathrm{SSD}}{dx}  =   \left\langle \sum_s\frac{d\theta_s}{dx}\right\rangle_t \, , \label{eq:dtheta_ssd}
  \end{align}
  where   $\langle X\rangle_t$ is the  average of $X$ over a $2\pi/\omega_m$ period (where $\omega_m$  is the main modulation frequency of the SSD device). The speckle bending rate, $d\theta_s/dx$, stems from Eq. \eqref{eq:bbf}.
Due to diffraction, $\sigma$ and $I_0$ respectively become $\sigma[1+(x-x_s)^2/z_c^2]^{1/2}$ and  $I_s/[1+(x-x_s)^2/z_c^2]^{3/2}$. The speckle reaches its maximum intensity $I_s$ at the position $x_s$, and has a Rayleigh length of  $z_c=\pi f_\sharp^2\lambda_0$, leading to 
  \begin{align}
 \frac{d\theta_\mathrm{SSD}}{dx}  &=   -S    \frac{n_0 }{n_c}  \frac{  I_0 }{ 2 v_g n_c T_e }    \frac{\langle \mathcal{G}  \rangle_\mathrm{\tau_\mathrm{SSD}}}{ \sigma}      \, ,  \label{eq:bbssd}
   \end{align}
 where $I_0$ now designates the beam averaged intensity and with 
   \begin{align}\label{eq:S}
   S &=\sum_s \frac{I_s/I_0}{[1+ {(x-x_s)^2}/{z_c^2}]^{3/2}}  \, . \nonumber \\
  \end{align}
The factor $S$, introduced in Eqs. \eqref{eq:bbssd}-\eqref{eq:S}, encompasses the speckle-dependent factors such as the speckle intensities, their spatial profiles and positions. As $S$ encloses the complex RPP-induced speckle dynamics,  we consider it as a parameter fitted from the paraxial simulations. This procedure further smooths out the hot spot contributions over the whole pulse.  This approximation implies to neglect the speckle disparities resulting from the variability of their intensity.
Furthermore, we neglect any significant modification of the speckle distribution during the beam propagation as it could affect the value of  $S$.
To mimic the effect of the SSD on the beam bending rate,  the speckles pattern is assumed to last a  time $\tau_\mathrm{SSD}$, proportional to the coherence time $T_\mathrm{SSD}$. The latter is  related to the modulation depth, $\Delta$ (defined hereafter at $3\omega$), and to the  modulation frequency of the SSD device, $\omega_m$, through $T_\mathrm{SSD}=2\pi/\omega_m (2\Delta +1)$. Introducing the factor, $\beta$, the parameter $\tau_\mathrm{SSD}$, referred as the speckle coherence time, writes,  
\begin{equation}
    \tau_\mathrm{SSD} = \beta \frac{2\pi}{\omega_m (2\Delta +1)} =\beta T_\mathrm{SSD}\, . \label{eq:taussd}
\end{equation}
This second factor, of the order of unity, represents the efficiency of SSD on the beam bending and  will be deduced from the paraxial simulations. 

Introducing the temporal mean of a function X(t) over a SSD period,  as
\begin{equation}\label{eq:avet}
    \langle X \rangle_\mathrm{\tau_ \mathrm{SSD}}= \frac{1}{\tau_\mathrm{SSD}} \int_0^{\tau_\mathrm{SSD}} X(t)dt \,,
\end{equation} 
Eq. \eqref{eq:G3} leads to,
\begin{align}
\langle  \mathcal{G}\rangle_\mathrm{\tau_\mathrm{SSD}} \simeq\frac{\sigma^3}{8\pi} \Im \int_{-\infty}^\infty dk  k^2  A_{k}e^{-k^2\sigma^2/4} \times\nonumber\\
\int_0^\pi d\theta \alpha_{k/f}[M_0\cos(\theta)]\cos(\theta) \times\nonumber\\
\left( 1+a_+\frac{ e^{g_+c_s\vert k\vert \tau_\mathrm{SSD}}-1}{g_+c_s\vert k\vert \tau_\mathrm{SSD}}- a_-\frac{e^{g_-c_s \vert k \vert \tau_\mathrm{SSD}}-1}{g_-c_s \vert k \vert \tau_\mathrm{SSD}} \right) \, .\label{eq:G3m}
\end{align} 
For sake of simplicity, we propose in appendix \ref{sec:fitg} a fully analytical fit of the above function, obtained in the fluid framework and valid in the regime
 $0.007 \le \gamma_0\le 0.05$ and  $\vert M_0\vert < 3$ for a mono-species plasma, illustrated in Figs. \ref{fig:fit}(a-d).

The mean deflection angle is related to the set of Eqs. \eqref{eq:dtheta_ssd}-\eqref{eq:bbssd}, and Eq. \eqref{eq:G3m}.
The dependence on the Mach number of the above predictions are illustrated in Fig.  \ref{fig:2d3dnl}(b)  for a He$^{2+}$ (blue lines), a C$^{6+}$ (red lines) and a Au$^{50+}$ (black lines) $10\%$-critical  plasma (with $T_e=2.5\, \rm keV$ and $T_i = 1\,\rm keV$) and for a speckle coherence time of $\tau_\mathrm{SSD}=7\,\rm ps$. The deflection rate is an increasing function of the ionisation number due to the thermal correction of Eq. \eqref{eq:nl}. In virtue of section \ref{sec:gauss3D}, the deflection rate is not peaked around $M_0=1$ owing to the short coherence compared to the transient.
A larger density causes the mean free path to decrease, thus increasing the local energy deposition and therefore enhancing the hole boring. 
The regime of validity of Eq. \eqref{eq:nl} ($\omega=kv_d<\nu_{ei}$) enclosed in the factor $\Omega$, may be assessed by estimating the critical Mach number $M_c = \nu_{ei}/[ c_sk _0/(2f_\sharp) ]$ above which the thermal correction is not valid anymore.  When $Z_i=2$ (we obtain $M_c \sim 0.7$), the factor $A_k$ is $\sim 20\%$ above its collisonless value  implying  a poor impact of thermal effects whatever the value of $\Omega$. However, the carbon deflection rates present slight inflections located around $M_0=M_c \simeq 1.7$, directly due to the condition $\omega=kv_d<\nu_{ei}$. 
Additionally, the  large value of $M_c \simeq 25$  obtained for the Au$^{50+}$ case  demonstrates the validity of the thermal corrections of Eq. \eqref{eq:nl} in this material.

%As expected, the 2D values (as blue lines) systematically overshoot the realistic geometry predictions. 
Because of the very large values of $Z_iT_e/T_i$ in the gold case  ($Z_iT_e/T_i=125$), the kinetic and fluid 3D deflection rates (as black and red lines respectively) concur \cite[]{POP_Ruyer_2020}.  For the same reason, accounting for an accurate distribution of gold ionisation levels at non-local-thermodynamic-equilibrium as predicted by DEDALE \cite[]{HEDP_Gilleron_2015} [in the multi-ion species formulation of Eq. \eqref{eq:drakek}],  only weakly reduces by $\sim 10\%$ the final bending rates compared to the mono-fluid average ionisation number model, as used in this study.

Finally, we performed five 3D hydrodynamic simulations with a paraxial solver (Hera~\cite{HERA_Jourdren_2005,Loiseau_2006}) of  Gaussian beam propagating in a C$^{6+}$ 10\% critical plasma with the laser and plasma parameters of Fig. \ref{fig:2d3dnl}(b) and for various Mach numbers. The simulation parameters and the method to measure the deflection rates are detailed in appendix \ref{sec:hera3d}. The thermal corrections of Eq. \eqref{eq:nl} (with $\Omega$ sets to 1) have been accounted for. The average over the first $7 \, \rm ps$ of the beam centroid deflection angles [$\langle\langle d\theta/dx \rangle_\perp \rangle_t$, Eqs. \eqref{eq:ave}, \eqref{eq:avet}] at the simulations exit plane are illustrated as red circles in Fig. \ref{fig:2d3dnl}(b) and  agree very well with the corresponding theoretical prediction (obtained with $S=\beta=1$) represented by the dashed red line. This confirms the validity of Eqs.  \eqref{eq:bbssd} and \eqref{eq:G3m} for a Gaussian beam. 
In the following,  the time averaged centroid deflection rate of a SSD beam will be related to the  above Gaussian speckle predictions by fitting the two scalar variables, $S$ and $\beta$, using dedicated 3D paraxial numerical results.

\section{Fit and validation of the bending rates by the means of paraxial hydrodynamic simulations }
\label{sec:parax}

We aim  at finalizing  and validating our speckle scale SSD beam bending model in light of paraxial 3D  %Hera \cite[]{HERA_Jourdren_2005, Loiseau_2006} and   
Parax  \cite[]{POP_Riazuelo_2000} hydrodynamic simulations. We will thus use in this section the fluid plasma response in our model [Eq. \eqref{eq:drakeh}]. Then, more realistic  predictions, such as in Sec. \ref{sec:icf}, will be made using the kinetic deflection rates. 
Ideally, the kinetic flow induced deflection should be fitted with kinetic  codes (such as particle-in-cell) -a  kind of simulation that is still out of reach with up to date supercomputers-.
In %both Hera and 
Parax code, the linear Landau damping operator is computed in the Fourier space \cite[]{Berger_98,POP_Rose_96,Masson_2006}   and the light is propagated through a paraxial solver.
Parax uses a linearized hydrodynamics module applied transversely to the main laser direction that will moderate the subsequent required numerical load.
This approach is consistent with the fact that density perturbations amplitudes accounted for in our model are always weak.
For sake of simplicity and comparison purposes, the bremsstrahlung energy deposition is also neglected. Note that as the pulse deflection is reduced by its loss of temporal coherence,  large simulation domains have to be considered in order to obtain clean quantitative comparisons.
The thermal corrections of Eq. \eqref{eq:nl} are %ignored in the 2D Hera simulations and 
accounted for in our 3D %Parax 
simulations.

%%%%%%%%%%%%%%%%%%%%%%%%%%%%%%%%%%%%%%%%%%%%%
%\subsection{Two dimensional HERA simulations }
%\label{sec:hera2d}
%
Parax has specific field injections and diagnostics to correctly simulate and characterize the different smoothing techniques. 
In order to reach more realistic ICF-conditions than in Ref. \cite[]{POP_Ruyer_2020}, the choice has been made to use a laser wavelength of $\lambda_0=0.35\, \rm\mu m$.
The focal spot is located at the center of the simulation domain, $x_\mathrm{foc}=1\, \rm mm$ with a focal number of $f_\sharp = 8.88$.
The  spectral dispersion  has a modulation frequency of  $\omega_m = 2\pi\times 14.25\,\rm GHz$.
The laser beam with $I_0=2\times 10^{14}\, \rm W/cm^2$ propagates through a $2\,\rm mm$ flowing plasma with $v_d/c_s= 0.9$ composed of Helium with $n_e=0.1n_c$, $T_e=2\,\rm keV$ and $T_i=500\, \rm eV$
while accounting for the thermal corrections of Eq \eqref{eq:nl}.
The simulations details are given in appendix \ref{sec:parax3d}. Different smoothing configurations have been simulated. 
\begin{figure}
\begin{tabular}{c}
\includegraphics[width=0.45\textwidth]{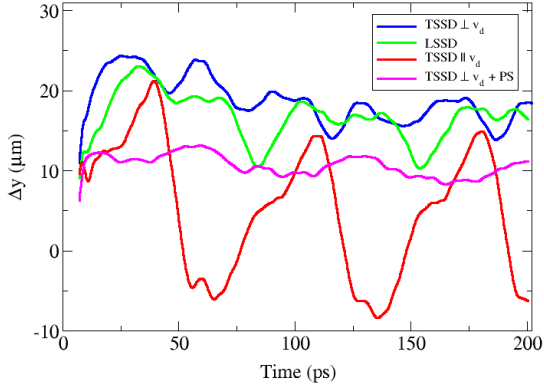}
\end{tabular}
\caption{ \label{fig:riazuello}
Temporal evolution of the beam centroid defined though Eq. \eqref{eq:ave} in the 3D Parax simulations with $\omega_m=2\pi\times 14.25\,\rm GHz$ and for $\Delta = 5.1$ for TSSD $\perp\mathbf{v}_d$, LSSD (green curve), TSSD $\parallel\mathbf{v}_d$ (red curve) and TSSD $\perp\mathbf{v}_d$+ PS (magenta curve).}
\end{figure}

Due to the phase modulation, the beam's center oscillates as soon as its injection, with a period corresponding to the modulation frequency, $2\pi/\omega_m\simeq 70\,\rm ps$. This effect leads to the centroid periodic motion at the simulation exit plane as shown in Figs.  \ref{fig:riazuello}. Note that the SSD beam bending model that we propose only predicts the temporal averaged of the beam centroid deflection, therefore neglecting the spatial envelope's motion.

%As expected, the increase of the  SSD modulation depth ($\Delta$ defined at $\lambda_0=0.35\,\rm \mu m$) decreases the beam deviation. 
\begin{figure}
%\begin{tabular}{cc}
%(a) 2D centroid displacement vs. $f_\sharp$  
%& (c) 3D paraxial simulations vs. theory \\
%\includegraphics[width=0.35\textwidth]{Fig3a.eps}
%& \multirow{3}{*}{
%\includegraphics[width=0.35\textwidth]{Fig3cnew.eps} 
%}
%\\
%(b) Impact of polarization smoothing&\\
%\includegraphics[width=0.35\textwidth]{Fig3psHe.eps}&
%\end{tabular}
\begin{tabular}{c}
(a) 2D centroid displacement vs. $f_\sharp$  \\
\includegraphics[width=0.35\textwidth]{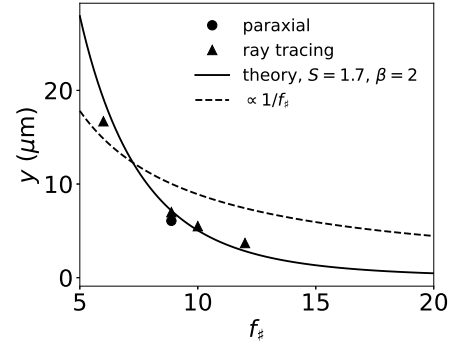}\\
(b) 3D paraxial simulations vs. theory \\
\includegraphics[width=0.35\textwidth]{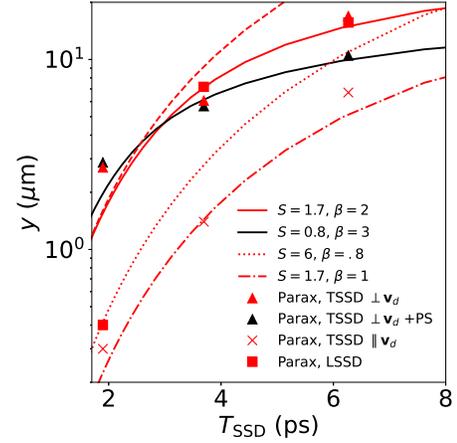} 
\end{tabular}
\caption{ \label{fig:ssd3d}
Centroid displacement after $2\,\rm mm$ of propagation as predicted by the theory as lines and by simulations as symbols. 
The plasma is composed of He$^{2+}$ with  $T_e=2\, \rm keV$, $Z_iT_e/T_i=8$,   $M_0=0.9$, $I_0=2\cdot 10^{14}\, \rm W/cm^2$. Panel (a) corresponds to a plot as a function of $f_\sharp$ with a fixed  $T_\mathrm{SSD}=3.7\,\rm ps$ (or a modulation depth of $\Delta=9$) while, in panel (b), $f_\sharp=8.88$ and $T_\mathrm{SSD}$ varies. 
(a) The predictions from the ray tracing scheme and from Eq. \eqref{eq:bbssd} with $S=1.7$ and $\beta=2$ correspond to the triangles and the black plain line respectively. A curve, $y\propto 1/f_\sharp$, is superimposed as a black dashed line and the 3D paraxial Parax simulation corresponds to the black circle. 
(b) The  predictions of Eq. (45) of Ref. \cite[]{POP_Rose_Ghosal_98} correspond to the red dashed line.
The case 
TSSD$\parallel \mathbf{v}_d$, TSSD$\perp \mathbf{v}_d$ and LSSD correspond to $(S,\beta)=(1.7,1)$, $(1.7,2)$, and   $(S,\beta)=(6.,0.8)$, respectively. The case with PS corresponds to $(S,\beta)=(0.8,3)$.
}
\end{figure}
The dependence of the beam's deflection angle against the f-number ($f_\sharp$) is illustrated in Fig. \ref{fig:ssd3d}(a). Note that, we varied the f-number, keeping constant the mean intensity, $I_0$. Within the anzatz of the  Gaussian beam deflection, the latter should be proportional to $f_\sharp$ (see the integration of a Gaussian beam bending deflection rate  over a Rayleigh length, Eq. (31) of Ref. \cite{POP_Ruyer_2020}). However, assuming $S$ may be replaced by a constant, Eq. \eqref{eq:bbssd} predicts $\theta_\mathrm{SSD}\propto \mathcal{G}\left(\frac{\tau_\mathrm{SSD}c_s}{f_\sharp \lambda_0}\right)/f_\sharp=o(1/f_\sharp)$ as illustrated by the plain line in Fig. \ref{fig:ssd3d}(a). Further, although  the speckle-scale contribution to the deflection increases with $f_\sharp$, the inter-speckle distance and  the transient regime amplitude both depend on $f_\sharp$. Hence, smoothed over the whole SSD beam, the resulting centroid deviation decreases faster than $1/f_\sharp$ [see dashed line of Fig. \ref{fig:ssd3d}(a)]. 
%For longer focal length ($f_\sharp=15$), the power of most of the speckles overcomes the self-focusing  threshold \cite[]{POP_Michel_2003} leading to the underestimate of the paraxial deflection by more than a factor of two (and  consistently with Fig. 2(b) of Ref.  \cite{POP_Ruyer_2020}), reminding us that the coupling of the beam bending with other instabilities such as filamentation or self focusing is out of reach of the present model.
 Figure \ref{fig:ssd3d}(a) also indicates that if the beam effective focal number $f_\sharp$ were to decrease, as expected when plasma smoothing occurs \cite[]{POP_Maximov_2001}, the resulting flow-induced deviation should be exacerbated.

%An additional set of four 2D paraxial Hera simulations has been performed in order to assess the impact of PS on the bending rates. Apart from the polarization smoothing added to the beam,  the same plasma and laser parameters as in Figs. \ref{fig:2d3dnl}(c,f) have been adopted and the centroid deviation, averaged over one modulation period (between $20$ and $90\,\rm ps$) after 2 millimeters of propagation is illustrated in Fig. \ref{fig:ssd3d}(b). It is shown that the use of PS slightly decreases the final beam deviation which, in our model corresponds to a reduction of the fitting factor $\beta$ from $8$ to $5$.

%Finally, the fair agreement obtained in a planar geometry between our simulations and theory over three different plasma parameters and four laser bandwidth validates the approximations behind our model. 
We may now fit our model with the most relevant smoothing techniques in light of paraxial simulations performed with the code Parax \cite[]{POP_Riazuelo_2000}. 
%\subsection{Three dimensional Parax simulations: SSD effects } \label{sec:fit3D}
%
In the case of  1D transverse smoothing by spectral dispersion (TSSD) oriented normally to the flow direction,  our numerical results represented by the red triangles in Fig. \ref{fig:ssd3d}(b), confirm that the increase of the SSD modulation depth (\emph{i.e.} the decrease of $T_\mathrm{SSD})$ lessens the amount of beam bending. Moreover, using $S=1.7$, $\beta=2$  (plain lines) correctly reproduces the centroid deviations.
Equation (45) of Ref. \cite[]{POP_Rose_Ghosal_98}, represented by the red dashed line, predicts the deflection of a beam assuming a Gaussian frequency spectrum. 
Its spectral width is thus narrower than for a SSD laser beam. Consequently, the beam bending dependence on the speckle coherence time is too steep and the predictions overestimate our numerical results for $\tau_\mathrm{SSD}\gtrsim 5\, \rm ps$.

When the TSSD direction is parallel to the flow, Ref. \cite[]{POP_Rose_Ghosal_98} has evidenced a significantly smaller deflection rate. The corresponding Parax simulations (as crosses) exhibit a deflection rate roughly $\sim 5$ times smaller than in the perpendicular configuration. In this case, the theoretical predictions with $S=1.7$ and $\beta=1$ give a good agreement (dotted dashed line). Those two cases confirm that for a beam smoothed by 1D transverse SSD, such as on NIF facility, different deviations are to be expected depending on the direction of the flow relative to the smoothing direction.
Indeed, when the direction of the spectral dispersion is perpendicular to the flow, the beam bending rate depends exclusively on the value of the flow. By contrast, when the direction of the spectral dispersion is parallel to the flow, the speckles are moving along the flow with super sonic velocities. The relevant parameter to assess beam bending is the difference between the flow velocity and the transverse velocity of the speckles, $\Delta v$. In TSSD, the time evolution of the speckle transverse velocity is roughly sinusoidal \cite[]{Videau_1999}. For example, for $\Delta=18$, the maximal transverse velocity reaches $v_y/c \sim 0.004$ which corresponds to $M\sim 3.3$. $\Delta v$ then evolves rather quickly as a function of time and is most of the time far from $c_s$, where the maximum of the deflection rate is. This explains both the small time averaged value of beam bending and the important centroid oscillations with time that can be seen in Fig. \ref{fig:riazuello} in which we represent the temporal evolution of the beam centroid for $\Delta = 5.1$ in the four simulated configurations for beam smoothing.

Additionally, the impact of PS has also been addressed for flows oriented normally to the transverse SSD direction. In our simulations, the polarizations in the near field are in the diagonal configuration as for the inner cones on the NIF facility \cite[]{NIF_user_guide}. 
Illustrated by  black triangles in Fig. \ref{fig:ssd3d}(b), our 3D Parax simulations fairly agree with our model  for  $(S,\beta)=(0.8,3)$.
They also suggest a weak  influence of the polarization smoothing on the bending rate, 
when the coherence time is small  ($T_\mathrm{SSD}< 4\,\rm ps$). For longer coherence time ($\Delta = 5.1$), beam bending appears to be reduced when PS is used in the simulation. This reduction can either be attributed to the uncorrelated spatial speckle distribution between the two polarizations or due to another non-linear effect. Indeed, when PS is not used, beam spreading is noticeable in the simulation (the angular aperture is increased by $\sim 20\%$, the effective $f_\sharp$ decreases from $9$ to $7.5$). The increase of beam bending without PS could be explained by the decrease of $f_\sharp$, as shown in Fig. \ref{fig:ssd3d}(a).
Given the weak influence of PS, the latter will be neglected when the flow is parallel to the transverse SSD direction.  

The dependence of the bending rate on the angle between the flow component normal to the main laser axis and the SSD direction, $\phi$, is out of the scope of the study.
Yet, the value of $\beta$ evolves from  $1$ to $2$ when $\phi=0$ (TSSD$\parallel\mathbf{v}_d$) and $\phi=\pi/2$ (TSSD$\perp\mathbf{v}_d$), respectively. The formula, $\beta(\phi) = 1+\vert \sin(\phi)\vert$, could be well suited to generalized the model to 3D.

Finally, another set of  Parax simulations were performed with  longitudinal temporal smoothing (LSSD), as used on the LMJ facility. In this configuration, the phase shift imposed between the different frequencies is radial, 
causing the speckles to move mainly longitudinally.
The results are represented by the red squares in Fig. \ref{fig:ssd3d}(c). The deflection rates are bounded by the two cases SSD$\perp\mathbf{v}_d$ and SSD$\parallel\mathbf{v}_d$ with the advantage to present a value independent of the flow direction. The use in our model of  $(S,\beta)=(6,0.8)$ leads to a fair agreement.

LSSD is more efficient than TSSD for small coherence time only. The LMJ optimum modulation depth is close to $\sim 15$ (defined for $\lambda_0=0.35\,\rm\mu m$) for $\omega_m=2\pi\times14.25\, \rm GHz$ leading to $\tau_\mathrm{SSD}\simeq 2.26\,\rm ps$ so that for the parameters of  Fig. \ref{fig:ssd3d}(b), the overall LMJ  deflection over 2 mm is about $\sim 1\, \rm \mu m$. 
The NIF currently uses $\tau_\mathrm{SSD}\simeq 6.5\,\rm ps$ ($\omega_m=2\pi\times17\, \rm GHz$  and $\Delta\sim4$) with PS and thus  corresponds to  a larger deflection than for the LMJ configuration ($\sim 10\,\rm \mu m$ and $\sim 3\,  \rm \mu m$ for a flow perpendicular and parallel to the TSSD direction, respectively). 
As a result, the beam bending mitigation  in high energy laser experiments is efficiently done by increasing the laser temporal spectral width. However, a large spectral width, apart from the associated FM-AM conversion effect that entails the laser performance \cite[]{Penninckx_FMAM,AO_Huang_2017},   may lead to anisotropic deflection rates when the TSSD is used. 

\begin{table}[]
    \centering
    \begin{tabular}{|c|c|c|}
             \hline
         & $S$ & $\beta$ \\
         \hline
         3D TSSD$\perp    \mathbf{v}_d$    & 1.7 & 2 \\
         3D TSSD$\perp    \mathbf{v}_d$+PS & 0.8 & 3 \\
         3D TSSD$\parallel\mathbf{v}_d$    & 1.7 & 1 \\
         3D TSSD$\parallel\mathbf{v}_d$+PS & 1.7 & 1 \\
         3D LSSD                           & 6. & 0.8 \\
         \hline
    \end{tabular}
    \caption{ \label{tab:bb}
    Values of the fitting parameter $S$ and $\beta$ to be used in Eq. \eqref{eq:bbssd}.}
\end{table}
Before including the present theoretical calculations in a ray tracing scheme, we gathered in table \ref{tab:bb} the values of the fitting parameters as obtained in the different geometries and SSD configurations.

\section{Accounting for beam bending with spectral dispersion in  ray tracing schemes}
\subsection{Description and validation of the ray tracing scheme}\label{sec:ray}
The time-averaged centroid deviation rate  may naturally be implemented in a ray tracing scheme as a correction to the well-known refraction effect issued from the Eikonal equation, by deflecting each ray accordingly to   Eq. \eqref{eq:bbssd}.
The ray direction and position ($\mathbf{k}$ and  $\mathbf{r}$ respectively)  therefore verify
\begin{align}
    \frac{d\mathbf{k}}{ds} &=-k_0 \frac{\nabla n_e }{2n_c\eta} -k_0   S    \frac{n_0 }{n_c}  \frac{  I_0 }{ 2 v_g n_c T_e }    \frac{\langle \mathcal{G} \rangle_\mathrm{\tau_\mathrm{SSD}}}{ \sigma} \frac{\mathbf{v}_\perp}{\vert \mathbf{v}_\perp \vert}  \, \nonumber\\
        \frac{d\mathbf{r}}{ds} &= \frac{\mathbf{k}v_g}{\omega_0} \, , \nonumber\\
    \mathbf{v}_\perp&= \mathbf{v}_d - \frac{ \mathbf{v}_d\cdot \mathbf{k}}{\vert  \mathbf{k}\vert^2}\mathbf{k}\, \label{eq:raybb}  
    %\mathbf{v}_\perp&= \mathbf{v}_d - \mathbf{v}_d\cdot \mathbf{e}_L\, \label{eq:raybb}
\end{align}
where $s$ is the ray abscissa, $\eta$ the optical index.

The above model has been implemented in a new  ray  tracing module of the hydrodynamic code Hera \cite[]{HERA_Jourdren_2005,Loiseau_2006}.
For solving the ray trajectory, we split the rectangular meshes of the 2D Hera hydrodynamic module  in four triangles using their diagonals.  In each triangle, the density profile is assumed to be linear allowing to have a continuous density profile in all the simulation domain. 
Hence, solving the Eikonal equation  in each triangle allows to split the ray trajectory in a sequence of parabolas, ideal for fast computing. 
Summing the contribution of each rays that pass through a given mesh following Ref. \cite[]{POP_Debayle_2019}, one obtains the local averaged intensity, $ I_0$, used in Eq. \eqref{eq:raybb}. To circumvent the non-linearity, we propose to use the local intensity calculated at the previous hydrodynamic time-step.

So has to mimic the averaged intensity profile at and off best focus, we define a lens position, hereafter at $x_\mathrm{lens}=-10$ m from which rays are shot. 
The rays departure $y$-positions  on the lens are chosen randomly, and their initial direction points toward a randomly distributed  position at the focal plane (at $x=x_\mathrm{foc}$)  following the measure given by the choosen intensity profile at best focus. 
Finally, the rays are propagated  in vacuum from the lens to the left boundary condition  of the simulation (at $x=x_\mathrm{BC}$).
In order to avoid statistical bias, the position and directions of the rays are shot at the beginning of each time steps. 
Note that this procedure, as used in Refs.  \cite[]{Lefebvre_2018}, prevents nonphysical thermal filamentation  arising from systematic statistical errors induced by the fluctuations of laser ray number per cell.
Until now, we assumed  $\mathbf{v}_d$ to be perpendicular to the main laser axis. For any flows, the axis along which the beam bending contribution in Eq. \eqref{eq:raybb}  lies results from the projection,  $\mathbf{v}_\perp$, of the fluid velocity $\mathbf{v}_d$ on the plane perpendicular to the beam main direction. 
Importantly, this modification to the classical ray tracing scheme [Eq. \eqref{eq:raybb}] remains minimal as it does not modify the  ray propagation main algorithm (described above) apart from adding a dependence on the beam intensity of the rays trajectories. 
Equation \eqref{eq:raybb} may   be applied in a 2D ray tracing simulation performed in the $(\mathbf{k}_0,\mathbf{v}_d)$-plane (where $\mathbf{k}_0$ is the main laser axis) with the 3D beam bending deflection rates [thus using  $  \langle  \mathcal{G} \rangle_\mathrm{\tau_\mathrm{SSD}}$, Eq. \eqref{eq:G3m} instead of  if its two dimensional equivalent]. Referred subsequently as a 2.5D geometry, this will allow us to estimate the bending rate in a realistic geometry in a cheap 2D simulation. 

In order to validate the implementation  of the beam bending model in the ray tracing module of Hera, we performed four 2.5D ray tracing Hera simulations. 
A mesh size of   $dx = 7.8\, \rm \mu m$ and   $dy =4\, \rm \mu m$ is used along with  $10^3$ rays with various values of $f_\sharp$. All other simulations characteristics (equation of states, boundary condition and simulation domain) are identical to the paraxial simulations, as described in Sec. \ref{sec:parax}.
Illustrated as black triangles in  Fig. \ref{fig:ssd3d}(a), the centroid deviations resulting from the modified ray tracing  module of Hera [using the fluid beam bending rate, Eqs. \eqref{eq:drakeh}, \eqref{eq:nl}, \eqref{eq:bbssd}  and \eqref{eq:G3m}], reproduce correctly the temporally averaged theoretical (plain lines) and paraxial (as a circle) predictions.

It is worth noting out that restricting the model to the  beam centroid deviation implies to neglect the  different deflections caused by  the various speckle intensities  and lifetimes,  contributing to the beam spreading. Likewise, the forward Brillouin instability \cite[]{POP_Grech_2006,PRL_Grech_2009,phd-Grech} and associated plasma smoothing effects may contribute to reduce the effective speckle waist thereby affecting the final beam centroid deviation \cite[]{POP_Maximov_2001}.
Figure \ref{fig:ssd3d}(a) demonstrates that the  bending rate decreases faster than $1/f_\sharp$, so that plasma smoothing effects may greatly worsen the impact of beam bending on the propagation of the laser.
Moreover, our crude modeling of the speckle dynamics does not  account for the oscillations of the coherence time and speckle velocity  as characterized in Ref. \cite[]{POP_Cain_Riazuelo_2012}, which also contributes to the fluctuation  of the beam direction.

\subsection{Quantifying the beam bending of a  realistic pulse in  ICF conditions}
\label{sec:icf}
\begin{figure}
\begin{tabular}{ccc}
(a) Total laser power \\
\includegraphics[width=0.4\textwidth]{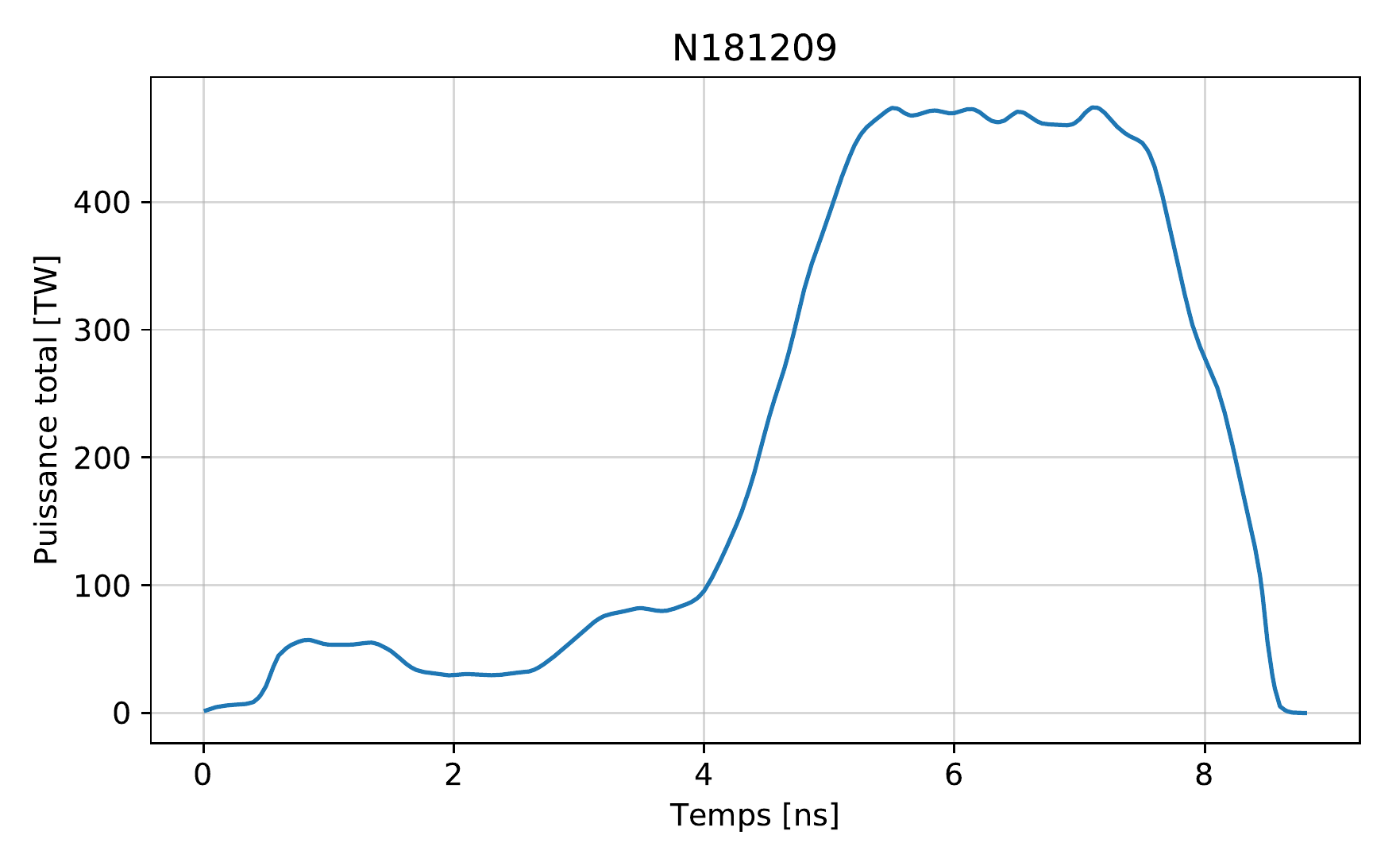}\\
 (b) $\langle Z_i\rangle $\\
\includegraphics[width=0.4\textwidth]{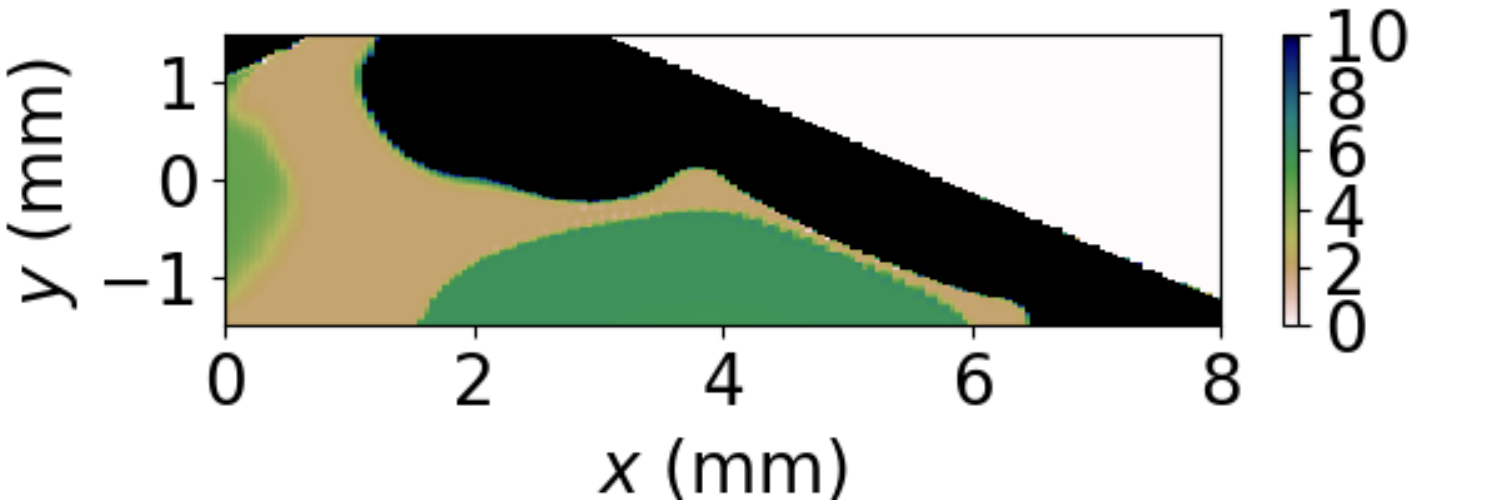}\\
(c) $T_e$ keV\\
\includegraphics[width=0.4\textwidth]{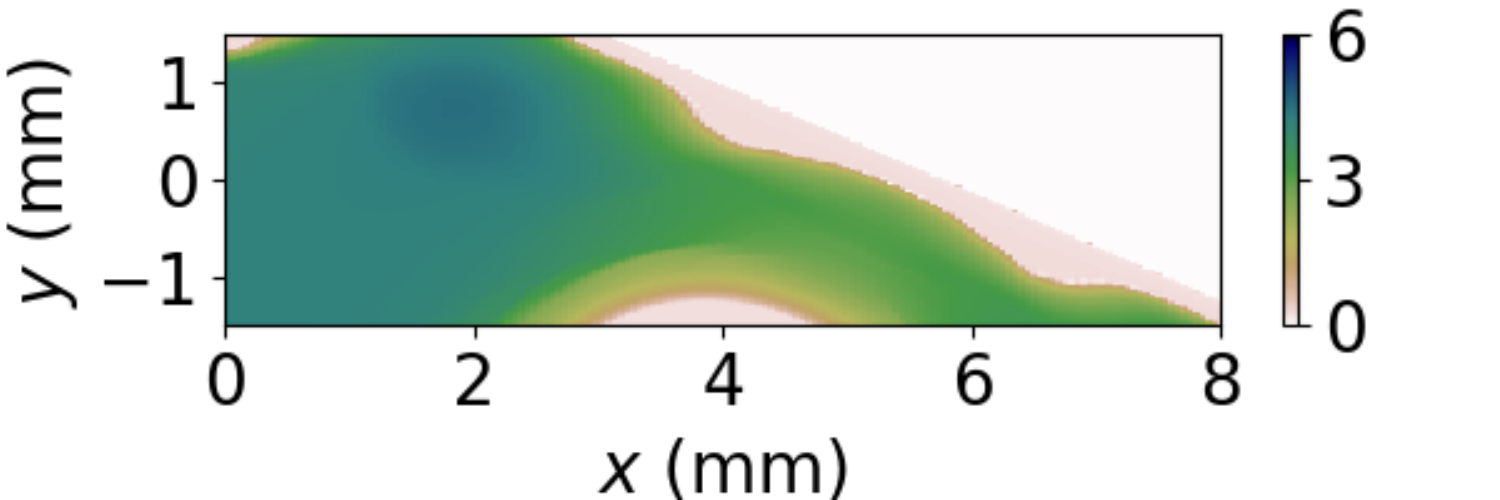}\\
(d) $n_e/n_c$\\
\includegraphics[width=0.4\textwidth]{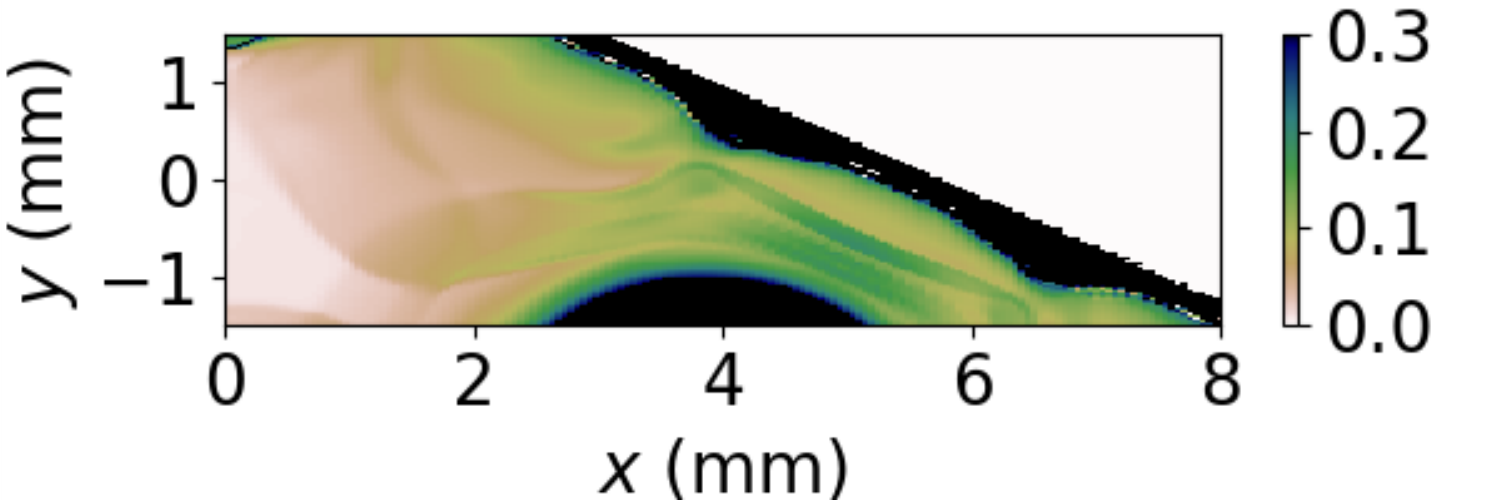} \\
(e) $\mathbf{v}_\perp /c_s$\\
\includegraphics[width=0.4\textwidth]{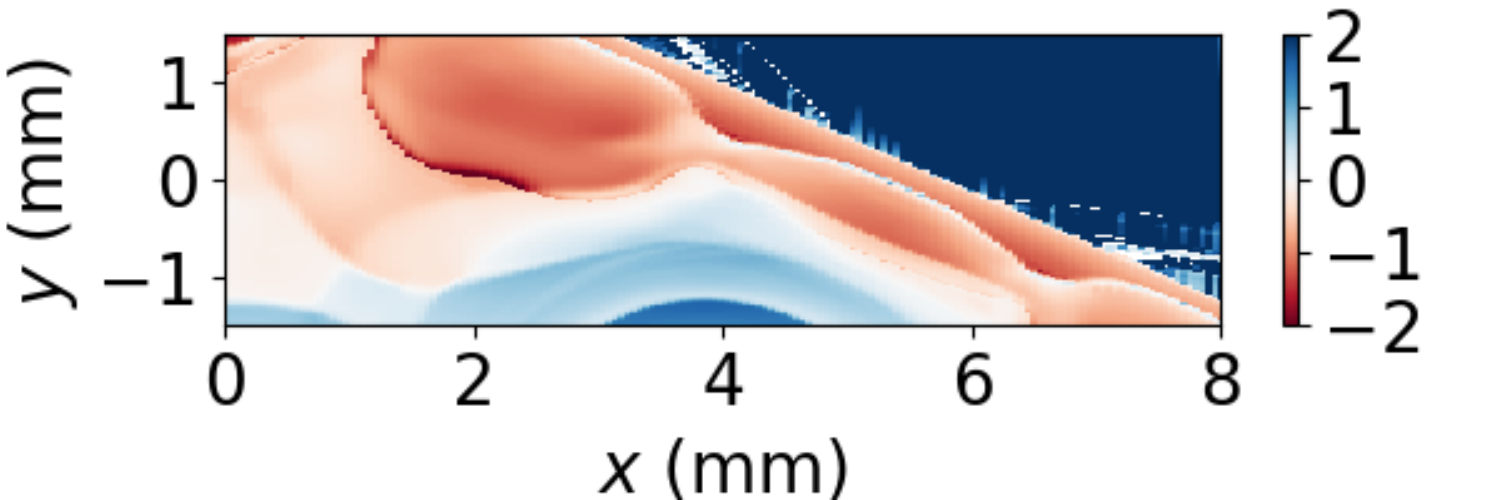} %\\
%(e) $d\theta_\mathrm{SSD}/dx$ ($^o/$mm) \\
%\includegraphics[width=0.4\textwidth]{dthetasdx6ns.png}\\
\end{tabular}
\caption{ \label{fig:icf} 
Hydrodynamic simulation performed with the code Troll \cite[]{Lefebvre_2018} for the  hybrid B NIF shot N181209 \cite[]{POP_Kritcher_2020,POP_Zylstra_2020,POP_Hohenberger_2020} 
%of a $1.8$ MJ NIF shot for a diamond ablator imploding capsule in a gold-lined hohlraum with low pressure gas-fill and 
illustrated in the frame of the inner cone ($y=0$ is the main inner cone axis), at 6 ns. 
(a) Total power of the main laser drive.
 Averaged local ionisation number, $\langle Z_i\rangle$ (colormap saturated to 10), b), electron temperature $ T_e$  (c), normalized electron density $n_e/n_c$ (colormap saturated to $n_e/n_c=0.3$), d) and Mach number of the $y$ flow velocity component (e).
}
\end{figure}
\begin{SCfigure*}
\begin{tabular}{cc}
(a) $\log(I\, [\mathrm{W/cm^2}])$, rays with bending &(b)  $\log(I\, [\mathrm{W/cm^2}])$, rays without bending\\
\includegraphics[width=0.33\textwidth]{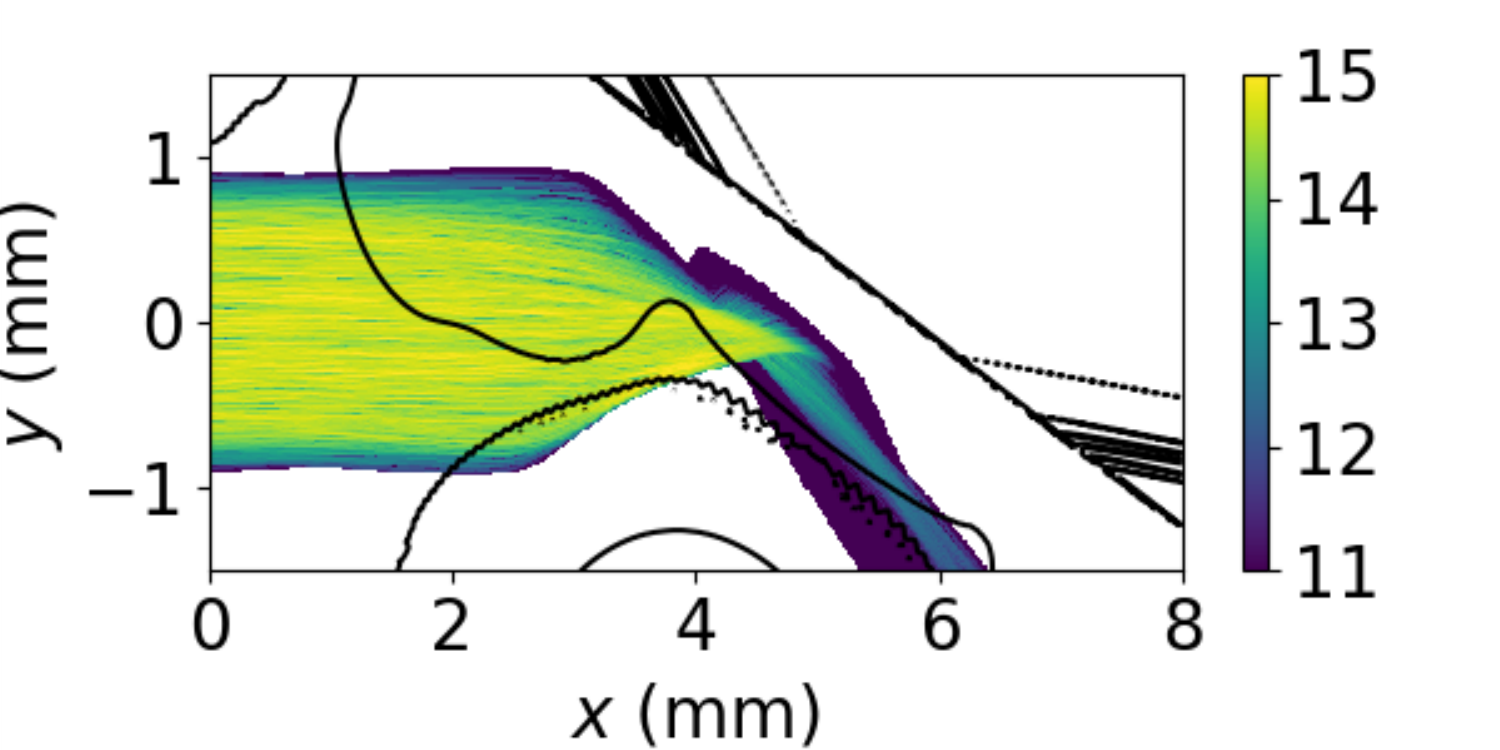} 
&\includegraphics[width=0.33\textwidth]{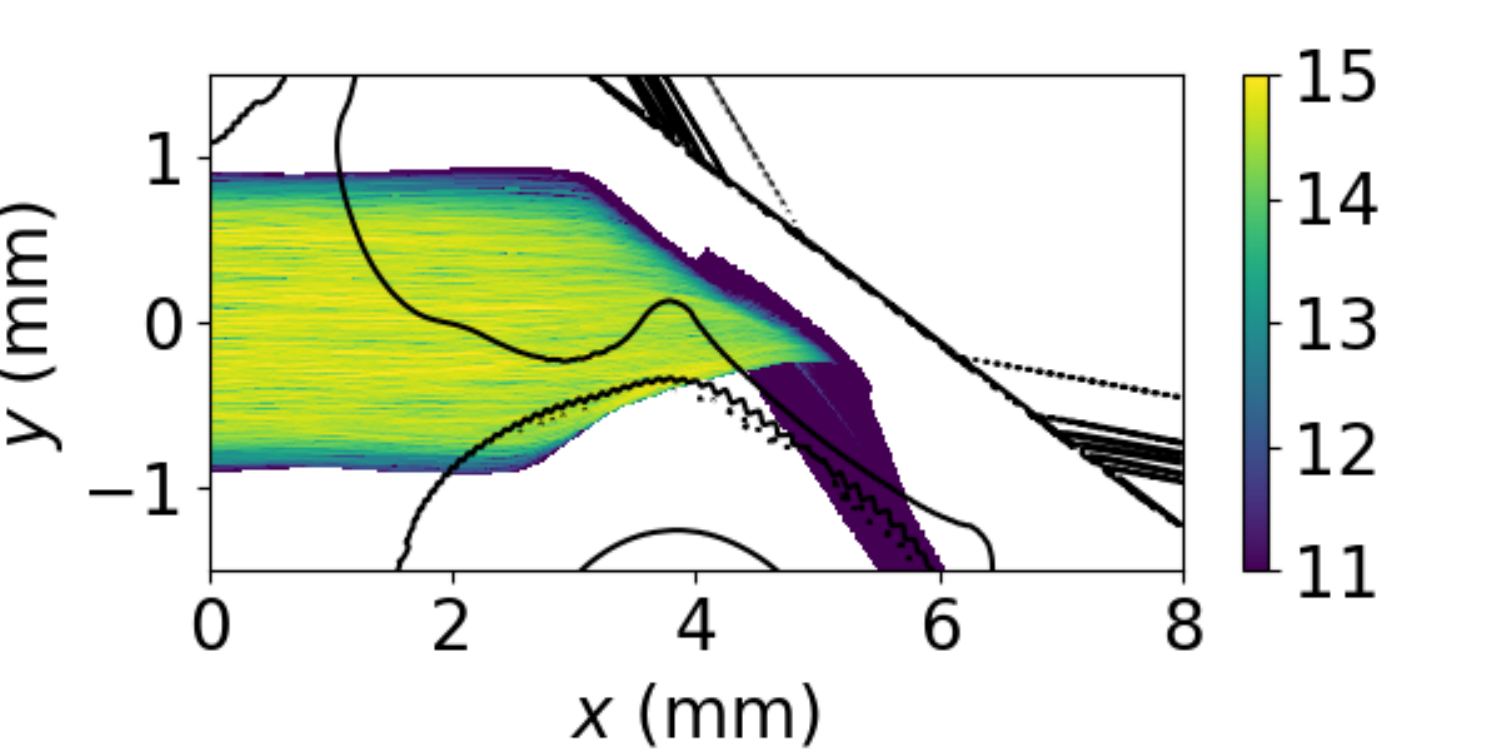} \\
(c) $S_{B}\, [\mathrm{W/mm^3}]$, rays with bending &(d)  $S_{B}\, [\mathrm{W/mm^3}]$, rays without bending\\
\includegraphics[width=0.33\textwidth]{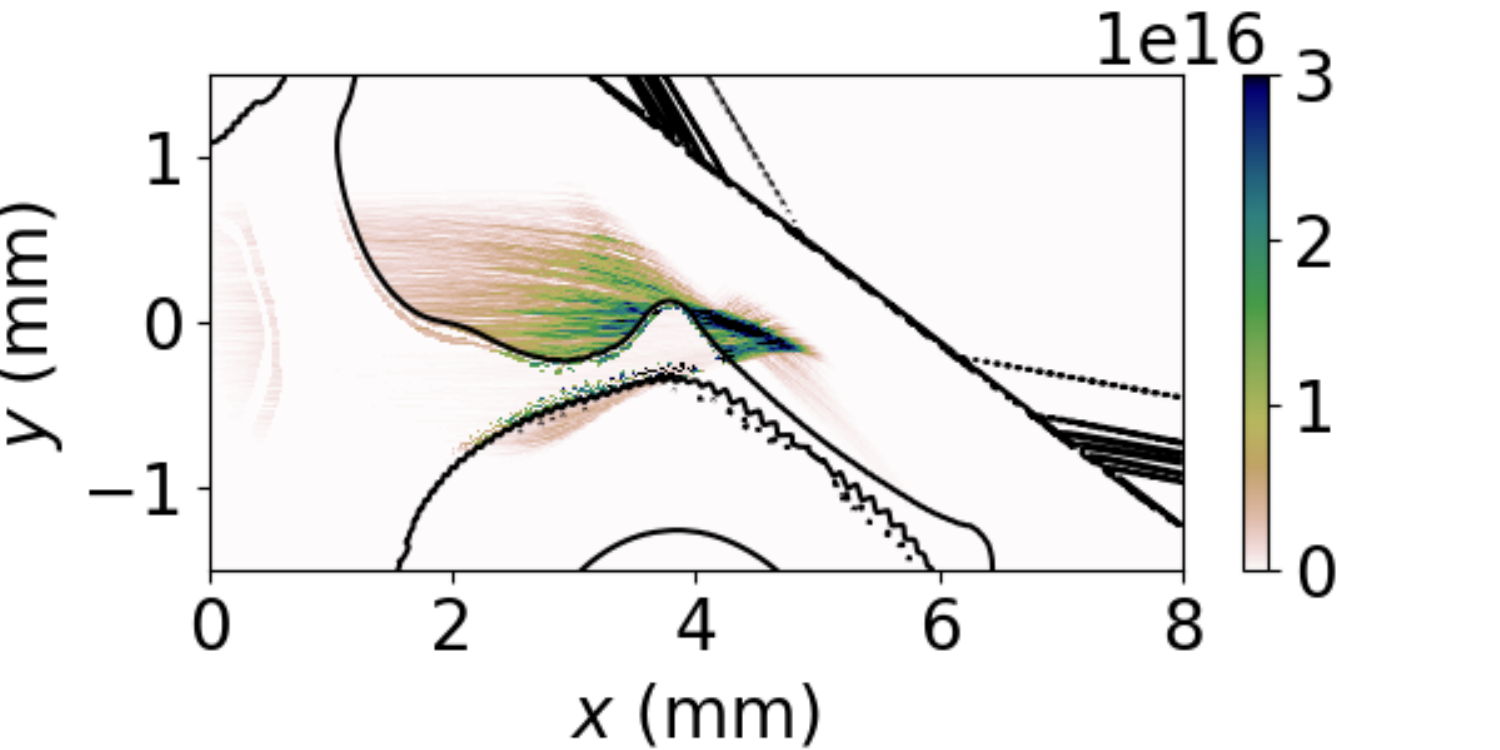} 
&\includegraphics[width=0.33\textwidth]{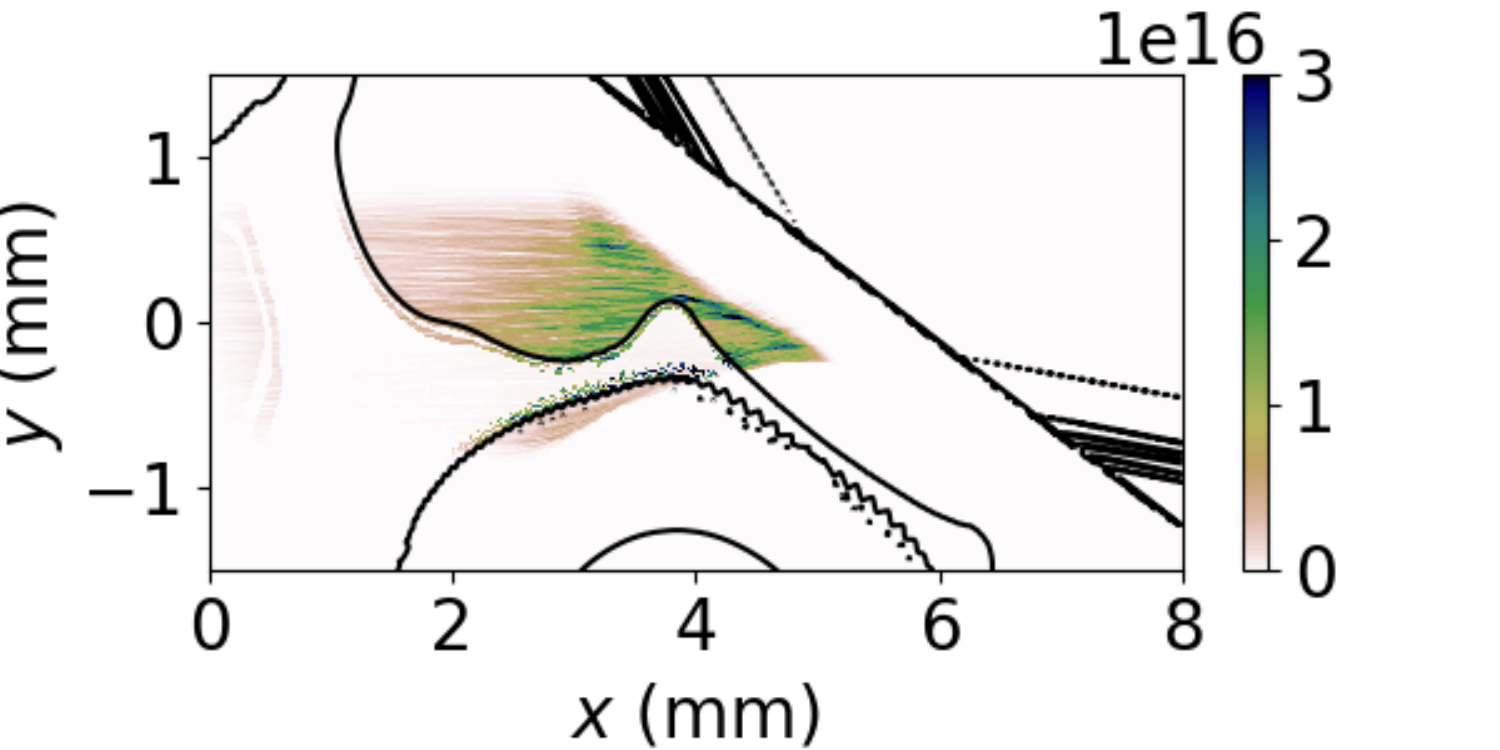}\\
(e) $d\theta_\mathrm{SSD}/dx$ (mrad/mm)&\\
\includegraphics[width=0.33\textwidth]{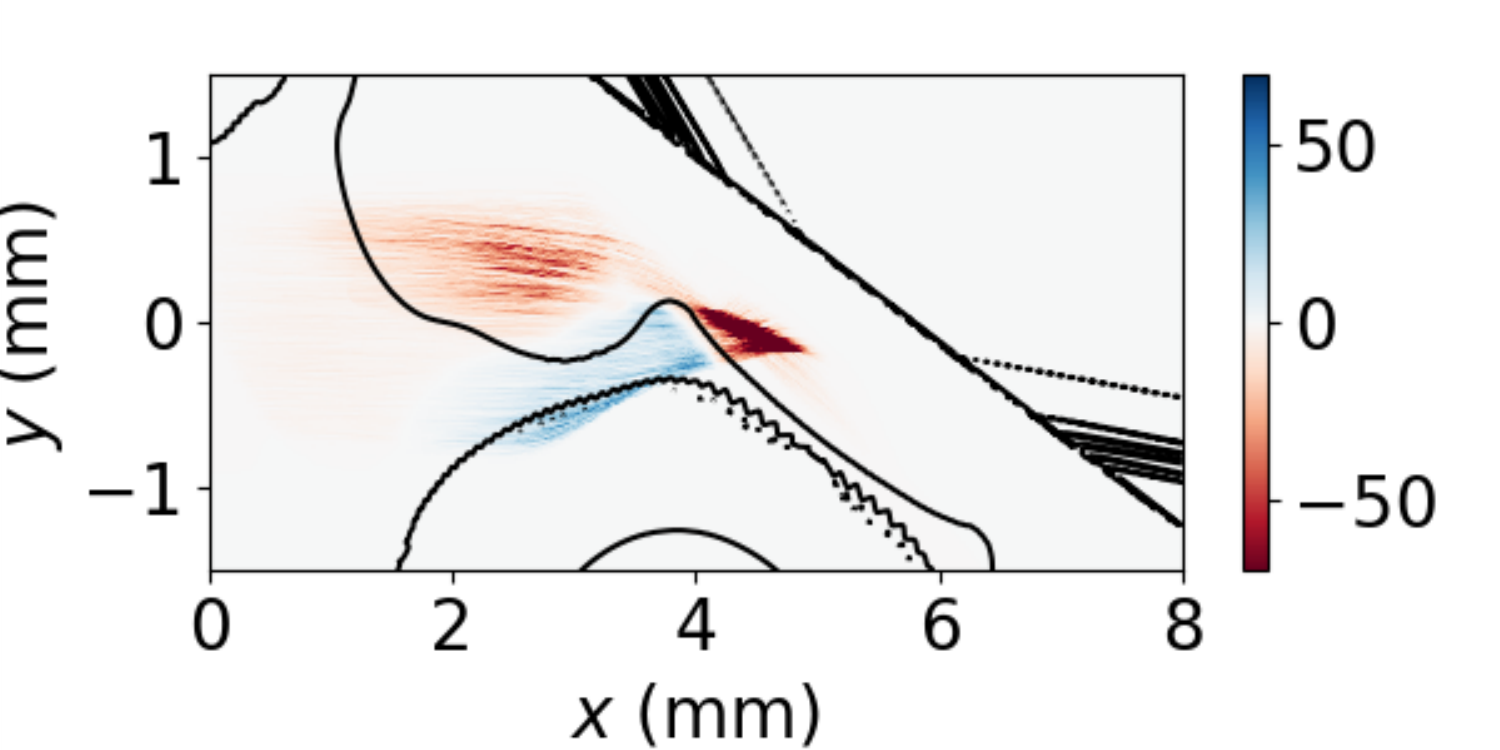} 
&
\end{tabular}
\caption{ \label{fig:icfbb} 
 Intensity profile (a,b) and deposited power  in the plasma through inverse bremsstrahlung ($S_{B}$, c,d) as predicted by the   2.5D beam bending rate [Eq. \eqref{eq:raybb} with $(S,\beta)=(1.7,1)$] (a,c,e) and classical (b,d) ray tracing scheme.
 (e) 2.5D deflection rate as predicted by Hera.
  The material boundaries are superimposed as black lines.  
}
\end{SCfigure*}
In order to quantify the beam bending level in realistic conditions, we performed a  2D-axisymmetrical Troll hydrodynamic simulation (see appendix \ref{sec:troll} for details) \cite[]{Lefebvre_2018}  of
hybrid B NIF shot N181209 \cite[]{POP_Kritcher_2020,POP_Zylstra_2020,POP_Hohenberger_2020}, \textit{i.e.} with 
a diamond  ablator and a low gas-fill hohlraum.
One  representative  time was retained,  $6\,\rm ns$, which corresponds  to  the  main power drive, as illustrated in Fig. \ref{fig:icf}(a).
We will here focus on the beam bending of the inner cone, whose main axis lies  on  $y=0$ and propagates  from left to right in  Figs. \ref{fig:icf}(b-e). 
The beam bending deflection rate is calculated using the local plasma and laser parameters ($f_\sharp=8$) with a speckle coherence time of $\tau_\mathrm{SSD}=7\,\rm ps$. Use is made of the 3D theoretical kinetic predictions of Eqs. \eqref{eq:drakek}, \eqref{eq:bbssd} and \eqref{eq:G3m} together with the nonlocal correction of Eq. \eqref{eq:nl}. 
Note that we do not use the analytical fit of $\langle \mathcal{G}  \rangle_\mathrm{\tau_\mathrm{SSD}}$ presented in  appendix \ref{sec:fitg} but we solve numerically the quadrature of Eq. \eqref{eq:G3m}. 
The  TSSD direction is vertical in the NIF chamber \cite[]{NIF_user_guide}, its direction therefore lies in the r-z plane of the hydrodynamic simulation. Owing to the weak effect of the PS, we will thus apply our deflection model with $(S,\beta)=(1.7,1)$, according to table \ref{tab:bb}.

Figures \ref{fig:icf}(b-e), correspond to the local $\langle Z_i\rangle$, $T_e$, $n_e/n_c$, $y$-flow velocity component, respectively. Note that the colormap is saturated to $\langle Z_i\rangle=10$,  thus the high-Z gold medium is represented by the black areas. The map of the y-velocity component, normalized to the local acoustic speed, suggests that, at $6\, \rm ns$, helium, carbon and gold, with the electronic densities ranging from $n_e\sim 0.05$ to $n_e\sim 0.2$, may be found on the inner beam path with $M_0\sim 1$. Indeed, as the helium gas is being compressed by the expanding  ablator [green region in Fig. \ref{fig:icf}(b)] and gold wall [black region in Fig. \ref{fig:icf}(b)],  the electron density remains in the $10\%$ critical density range. Likewise, the electron temperature, higher than $3$ keV in these regions, leads to qualitatively small Landau damping rates and  therefore to a bending not so far from its large asymptotic limit. 
Note that, at $6\,\rm  ns$, the density of the expanded window is of the order of $n_e\sim 10^{-3}n_c$  [see Fig. \ref{fig:icf}(b)]  which is too small for the speckle scale beam bending to occur.

Two 2.5D ray tracing HERA simulations have been performed using the plasma profiles resulting from the Troll simulation at $6\, \rm ns$. 
These simulations do not solve the hydrodynamic equations. 
We use a domain size of $L_x\times L_y = 8 \times 3 \, \rm mm^2$, a mesh size of  $dx = 9.7\, \rm \mu m$ and $dy =11.7\, \rm \mu m$, a focus of $f_\sharp=8$ and a beam waist of 1 mm.
The beam is composed of $10^3$ rays, all other simulations characteristics are identical to the paraxial simulations.  
Figures  \ref{fig:icfbb}(a,c) present the intensity map and the inverse bremsstrahlung power of the inner cone with the modified scheme of Eq. \eqref{eq:raybb}. Both figures are to be compared with the  classical ray propagation (b,d). The resulting intensity profiles  evidence a weak impact of the beam bending dynamics on the pointing direction.  When accounting for the beam deviation (a), the expanding gold bubble around  $x=2\,\rm mm$ tends to deflect the laser toward the ablator. Indeed, Fig. \ref{fig:icfbb}(e) illustrates the local 2.5D bending rate and shows a slight deflection of $\sim -25 \, \rm mrad/mm$ located in the gold material. Likewise, the compressed helium and a small part of the ablator tends to deviate the other half of the beam   (by $\sim + 25 \, \rm mrad/mm$)  in the direction of the gold wall (upward) thus enhancing the already present focusing effect due to refraction.
The temporal incoherence of the laser is thus not strong enough to totally suppress the beam bending and compensate the thermal effects, especially in the gold medium, leading to significant deflection rates, as illustrated in Fig. \ref{fig:icfbb}(e). 
Quantitatively, the energy deposition is $0.5\%$ weaker in the gold medium as extracted from our ray tracing simulation  with our beam bending model than without and $10\%$ less power is deposited onto the carbon ablator when the deviation is accounted for.
However, the profile of the laser absorption is  modified as it is much more peaked   around $(x,y)\simeq (4.5,0)\, \rm mm$ when the  beam bending is included.
Quantitatively,  $\sim 600\, \rm GW$ of laser power is able to leave the gold bubble with beam bending, which greatly contrasts  with the $\sim 20\, \rm GW$ obtained without flow-induced refraction. These numbers, highly sensitive to the beam bending model, seem to be at least qualitatively consistent with the measurements
presented in  Ref. \cite{abstract_Lemos_Glint}.

Importantly, the part of the beam that is specularly refracted by the gold wall is more energetic with  our model  than with the classical ray tracing scheme [see Figs. \ref{fig:icfbb}(a,b) around $(x,y)\simeq (6,-1)\, \rm mm$]. The small flow-induced deflections in the gold bubble seem to be sufficient, at grating incidence, to significantly modify the penetration depth of the laser in the dense plasma and its absorption. 
The resulting deflected beam, of intensity in excess of $10^{13}\,\rm W/cm^2$, could reach the opposite laser entrance hall 
and therefore  could lead to a possible wave mixing with opposite-side beams. 
The striking qualitative similarity with experimental measurements ascribed to stimulated Brillouin side scatter or the so-called glint \cite[]{Honda_1998,PRL_Turnbull_2015,abstract_Lemos_Glint}  urges to address the impact of beam bending on  the implosion dynamics in ICF experiments.

\section{Conclusions and prospects}
Based on a previous publication, we addressed the beam bending physics of a realistic (RPP, SSD and PS) high-energy laser beam  propagating in  ICF plasma conditions. 
By mean of large homogeneous paraxial hydrodynamic simulations, we constructed a 3D ray-tracing model which  quantitatively captures the temporally averaged  centroid deviation  of realistic  beams.
The use of our ray tracing model in ICF hybrid B conditions shows a weak modification of the beam propagation properties  resulting from the speckle scale beam bending. Our results  indicate that the flow-induced deviation in the gold bubble is a good  candidate for explaining the anomalous refraction measurements, the so-called  glint \cite[]{Honda_1998,PRL_Turnbull_2015}. Our model may thus be used to predict and assess the impact of this anomalous refraction on the implosion dynamics and wave mixing processes. 
The impact of beam bending in other ICF scenarios is left for future work. 

Hence, the speckle-scale physics  may be included in a coarse description of light such as  ray tracing-based schemes, opening the way for a more realistic laser energy deposition model in radiative hydrodynamic codes.
Achieving more accurate predictions of the implosion symmetry and bang time without resorting to non-physical artefacts such as laser power multipliers \cite[]{POP_Kritcher_2018} may allow   smart ICF designs \cite[]{POP_Vandenboomgaerde_2018,PRL_Depierreux_2020} and a possible control of deleterious laser plasma effects.

Additionally, our ray tracing model with spectral dispersion has been compared to   homogeneous ideal simulations, so that, its   validity should be ensured in   smooth density gradients only. 
Furthermore, most wave-mixing processes are absent from our analysis, possibly affecting the beam deflection and bringing further complexity to the system \cite[]{POP_Huller_2020}. 
According to Fig. \ref{fig:ssd3d}(a), the speckle scale beam bending may  be greatly amplified by plasma smoothing effects \cite[]{POP_Maximov_2001,POP_Yahia_2015}. 
The ICF estimate of Figs. \ref{fig:icfbb}(a,c,e) could thus represents a lower bound of the realistic flow-induced deflection.
Since our study is limited to the beam centroid deviation, further analysis, including forward scattering and beam spreading \cite[]{PRL_Grech_2009,POP_Hinkel_1998,PRL_Turnbull_2022}, are required in order to properly understand all the physical mechanisms able to affect the beam pointing and subsequent laser energy deposition region inside the hohlraum.

Furthermore, our theory and simulations in realistic geometries allow to compare the efficiency of the various temporal smoothing techniques  in order to reach control of the beam pointing direction. Compared to LSSD, we confirm that the 1D TSSD results in an anisotropic deflection rate. The latter can be substantially reduced by increasing the SSD modulation depth to reach field coherence time below 2 ps. 

Finally, a ray tracing-based  simulation that would include our  beam bending model requires to evaluate the quadrature of Eq. \eqref{eq:G3m} each time a ray exits a cell. The exact calculations seem  numerically too constraining unless one proceed to a tabulation of the deflection rates or to an accurate fit that would significantly alleviate the computational  load. 
For that purpose, a fit is presented in appendix \ref{sec:fitg} which should help an implementation of a ray tracing beam bending model in  hydrodynamic codes.

\section*{Data availability}
The data that support the findings of this study are available from the corresponding authors upon reasonable request.

\section*{Acknowledgements}
We acknowledge fruitful discussions with S. Laffite, D. Hinkel,  W. Farmer and important inputs from D. Penninckx and J. M. Di Nicola. We also acknowledge F. Gilleron and R. Piron for the use of the DEDALE model and D. Dureau and O. Morice for their advices on the ray tracing algorithm. We also acknowledge the HERA team for the developments and maintenance of the simulation code.
This work has been done under  the auspices of  CEA-DAM and
the simulations were performed using HPC resources at TGCC/CCRT and CEA-DAM/TERA.

\appendix

%\section{Appendixes}
\section{Three dimensional Hera simulation of the Gaussian beam bending}
\label{sec:hera3d}
\begin{figure}
\begin{tabular}{cc}
\includegraphics[width=0.24\textwidth]{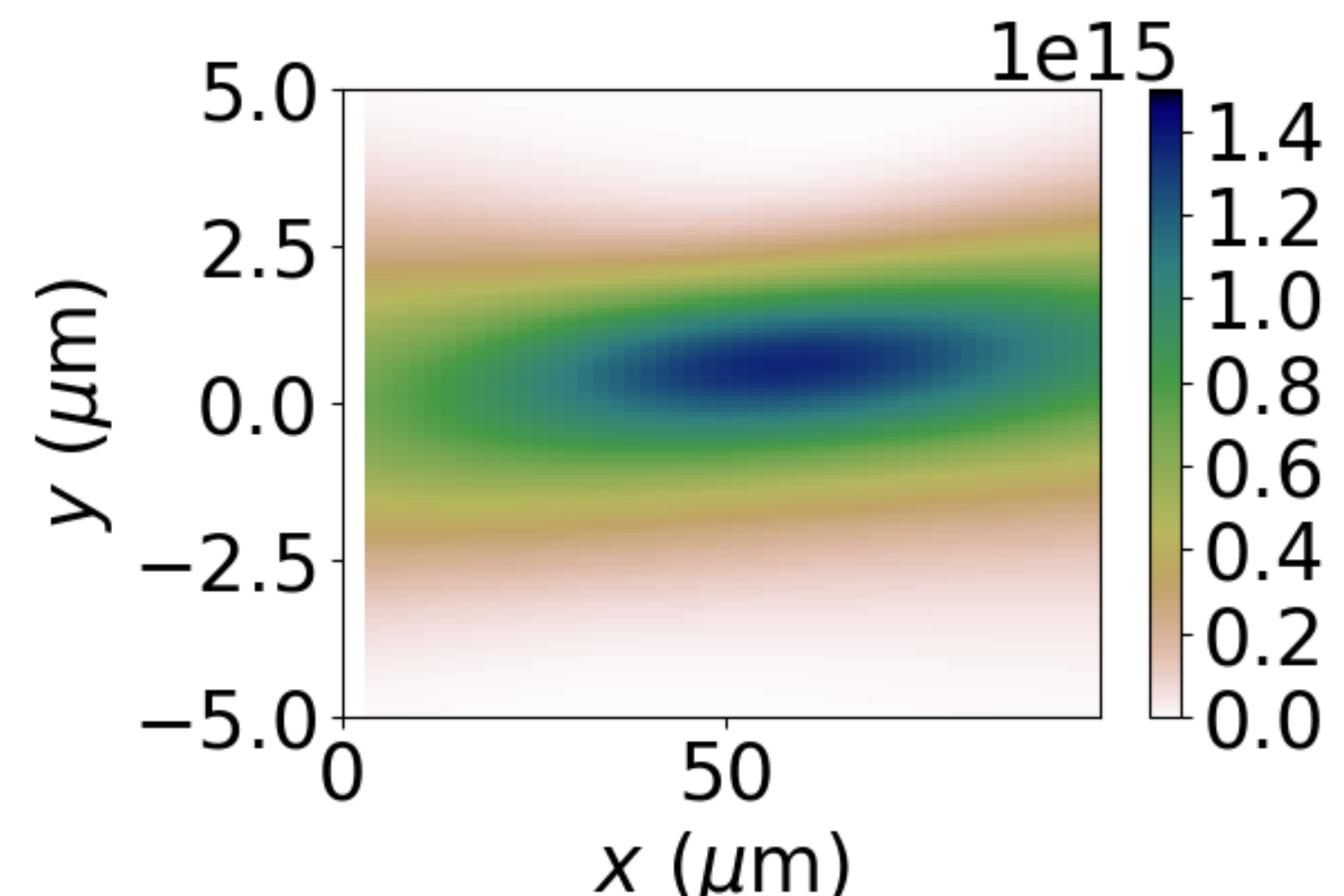} &\includegraphics[width=0.24\textwidth]{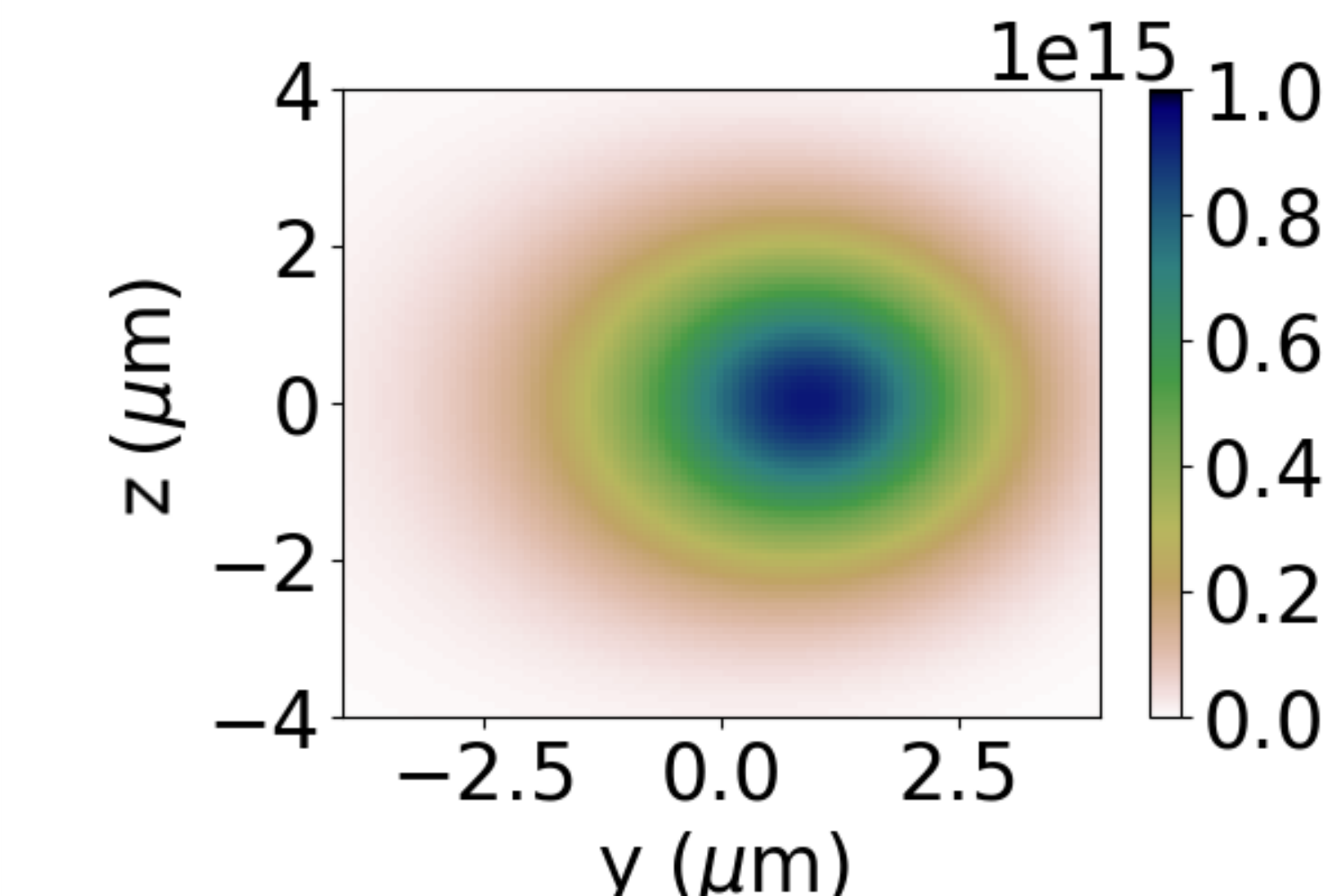}
\end{tabular}
\caption{ \label{fig:hera3d} 
Intensity profiles [$\rm W/cm^2$] of a 3D Gaussian laser pulse propagating through a flowing ($M_0=0.8$) fully ionized carbon plasma at 10\% to the critical density  as predicted by 3D  Hera simulations.
(left) Intensity profile in the  plane $(x,y)$ with the flow velocity along the $y$-axis and the main laser direction along the $x$-axis. (right)  Intensity profile in the exit plane. 
}
\end{figure}
For validating the predictions of Sec. \ref{sec:gaussssd},
five three dimensional Hera simulations of Gaussian beams  have been performed with a wavelength of $\lambda_0=0.35\,\rm\mu m$, a maximum intensity of $I_0=10^{15}\, \rm  W/cm^2$ and a focal number of $f_\sharp = 8$. 
The homogeneous fully ionized carbon plasma at 10\% to the critical density has a drift velocity along the $y$-direction at a Mack number of $M_0 = 0.4$, $0.8$, $1$, $1.2$, or $1.5$ with $(T_e, T_i) =(2.5 ,1) \, \rm keV$. The size of the simulation domain is $L_x\times L_y \times L_z = 100 \, \mathrm{\mu m} \times ( 40 \, \mathrm{\mu m} )^2$ and is composed of  $50\times512^2$ meshes. The laser is injected from the left boundary at $x=0$  and its focal spot is located at $(x_\mathrm{foc},y_\mathrm{foc},z_\mathrm{foc}) = (50\,\mathrm{\mu m}, 0, 0)$. 
The laser has a constant temporal profile preceded by a $1\, \rm ps$-long linear rise. 
Additionally, the Landau damping operator is calculated in the Fourier space transversely to the main laser direction  \cite[]{Berger_98,POP_Rose_96,Masson_2006} and the thermal correction of Eq. \eqref{eq:nl} (while setting $\Omega=1$) is accounted for.

The resulting intensity profiles at $t=10\, \rm ps$, illustrated in Fig. \ref{fig:hera3d}, unravel a deviation of the beam toward the flow ($y$) direction. 
The theoretical predictions of Sec.  \ref{sec:gaussssd}, that we aim at validating, gather all the $x$-dependence of the bending rate in the fitting factor, $S$. Hence the comparison of the 3D Hera predictions with Eq. \eqref{eq:bbssd} requires to remove from the simulations results the influence of the $x$-Gaussian laser profile. 
We thus start by isolating the $x$-dependent factor (waist and intensity), leading in 3D to 
\begin{equation}
    \frac{d \theta}{ dx} \simeq \frac{d^2y}{dx^2} = \frac{1}{(1+(x-x_\mathrm{foc})^2/z_c^2)^{3/2}} \frac{d\theta_0 }{dx} \, ,
\end{equation}
where $z_c=\pi f_\sharp^2 \lambda_0$ and $d\theta_0 /dx $  is the part of the beam bending rate that is independent of $x$, thus corresponding to Eq. \eqref{eq:bbf}.
Hence, for $d\theta/dx(x=0) = y(x=0)=0$, the deviation may be related to $d\theta_0 /dx $ through 
\begin{equation}
    y \simeq \frac{1}{(1+(L_x-x_\mathrm{foc})^2/z_c^2)^{1/2}} \frac{L_x^2}{2} \frac{d\theta_0 }{dx}\, .
\end{equation}
The value $S=1$, imposed in   Sec.  \ref{sec:gaussssd},  corresponds to a deviation that reads $y_0 = 0.5L_x^2d\theta_0 /dx$.
In summary, in order to extract the relevant time averaged centroid deviation from our 3D Hera simulation, we first compute the centroid deviation, $\langle y \rangle_\perp$ at the exit plane [using Eq. \eqref{eq:ave}], divide it by the factor  $(1+(L_x-x_\mathrm{foc})^2/z_c^2)^{-1/2}\simeq 0.81$ and averaged the obtain results between $t_0$  and $t_0+7\, \rm ps$. We use here  $t_0=1.3 \, \rm ps$  to account for the time required for the laser to reach the simulations exit plane ($\sim 0.3\,\rm ps$) and the $1\, \rm ps$ linear rise of the laser time envelop. 

\section{Three dimensional Parax simulations}
\label{sec:parax3d}
\setcounter{equation}{0} 
\renewcommand{\theequation}{B\arabic{equation}}
The code Parax   simulates the propagation of electromagnetic waves in a plasma \cite[]{POP_Michel_2003}. 
%We shall denote by z the main propagation axis, x, y the two transverse axes and t the time, respectively. 
The propagation of one fixed polarization electromagnetic wave is modeled by one generalized scalar paraxial equation \eqref{eq:parax1} for the electric field amplitude $E$ and a wave equation \eqref{eq:parax2}  for the plasma response in the perpendicular ($y,z$)-plane:
\begin{align}
    \Big(&  2i\frac{\omega_0}{c^2}\partial_t  +2ik_0\eta \partial_x +i \partial_x k_0\eta+ \nonumber\\& \nabla_\perp^2-\frac{\omega_0^2}{c^2}\frac{n_e-n_{e0}}{n_c} -\frac{\nu_{ei}\omega_0n_{e0}}{c^2n_c} \Big) E = 0 \, . \label{eq:parax1}
\end{align}
Here, $\nu_{ei}$ is the electron–ion collision frequency. Stimulated Raman and Brillouin backscattering do not occur due to the presence of a single paraxial incident wave.  However, both filamentation and FSBS can grow and interact.  
The plasma density is modeled using a fluid description, where expansion to second order in the field perturbation leads to an ion-acoustic wave driven by the ponderomotive and thermal effects \cite[]{POP_Michel_2003,Walraet_2003}:
\begin{equation}\label{eq:parax2}
    \left[  (\partial_t+\nu +v_y \partial_y )^2 -c_s^2 \nabla_\perp^2 \right]\log\left(\frac{n_e}{n_{e0}}\right) = \frac{Z_i}{cm_in_c}\nabla_\perp^2( A_t I ) \, .
\end{equation}
Here, $v_y$ stands for a transverse plasma drift assumed to be along the $y$-axis and $\nu$ is the ion acoustic wave damping rate.
The logarithm saturates the density response and so impedes the blowup of the self-focusing process that would otherwise be induced by the cubically nonlinear Schr\"odinger equation derived from Eq. \eqref{eq:parax1}. 
Equation \eqref{eq:parax2}  uses the acoustic type of plasma response and accounts for the plasma heating using a nonlocal electron transport model according to \cite[]{POP_Brantov_1998}. The operator $A_t$ in the source term is applied to the laser intensity; it accounts for the inverse bremsstrahlung heating, the ponderomotive effect, and the nonlocal transport. Its spectrum in the Fourier space for the transverse spatial coordinates $y$, $z$ uses the fit as introduced in Ref.  \cite[]{POP_Brantov_1998}.

In the case of PS smoothing, the code models the propagation of two independent electromagnetic waves (the first one polarized along the y-axis, the second one along the z-axis) but the source term of (B2) is computed with the contribution of both electromagnetic waves.

Equation \eqref{eq:parax2} has been derived using  the small amplitude density and temperature hypothesis : $dn_e/n_e \ll1$, $dT_e/T_e \ll 1$  and the  quasi-stationary temperature hypothesis: $dt(T_e) <\nu_{ei} I_0/cn_c$.
In our simulations, the laser propagates into a 2 mm Helium plasma which is assumed to be uniform with density $n_e(t=0) =0.1n_c$. $T_e=2\, \rm keV$ and $T_i = 0.5\, \rm keV$. The Landau damping rate has been determined by a kinetic dispersion solver and its value is $\gamma_0=0.031$. The laser averaged intensity is $\langle I_0\rangle= 2\times 10^{14}\, \rm W/cm^2$. This relatively small value has been chosen in order to limit the angular spreading of the beam. The transverse and longitudinal mesh sizes are $dy = dz = 0.75\lambda_0$ (for $\lambda_0.35\, \rm \mu m$) and $dx= 1.6 \, \rm \mu m$, respectively.

\section{Troll 2D axisymmetrical simulation}
\label{sec:troll}
\setcounter{equation}{0} 
\renewcommand{\theequation}{C\arabic{equation}}
Troll is a 3D Arbitrary Lagragian-Eulerian (ALE) radiative hydrodynamic code with unstructured mesh \cite[]{Lefebvre_2018}. The simulation presented in Sec. \ref{sec:icf} was performed in 2D axisymmetrical geometry. We used the latest CEA’s tabulated equations of state and opacities, with a model for NLTE correction on the emissivity. The radiation transport is solved using an implicit Monte-Carlo method and the heat flux using the Spitzer-Härm model, limited to 10\% of the free streaming flux. The lasers are simulated using the classical 3D ray tracing method (without beam bending model), with inverse bremsstrahlung absorption, corrected to account for the Langdon effect. Both the Raman and Brillouin backscattering were measured to be a few percents, and thus, not accounted for. No multipliers were used on the incident laser power.

\section{Fit of Eq. \eqref{eq:G3m} in the fluid framework}
\label{sec:fitg}
\setcounter{equation}{0} 
\renewcommand{\theequation}{D\arabic{equation}}
\begin{figure*}
\begin{tabular}{cc}
(a) $\gamma_0=0.007$, Au$^{50+}$  & (b)  $\gamma_0=0.02$, Au$^{50+}$    \\
\includegraphics[width=0.4\textwidth]{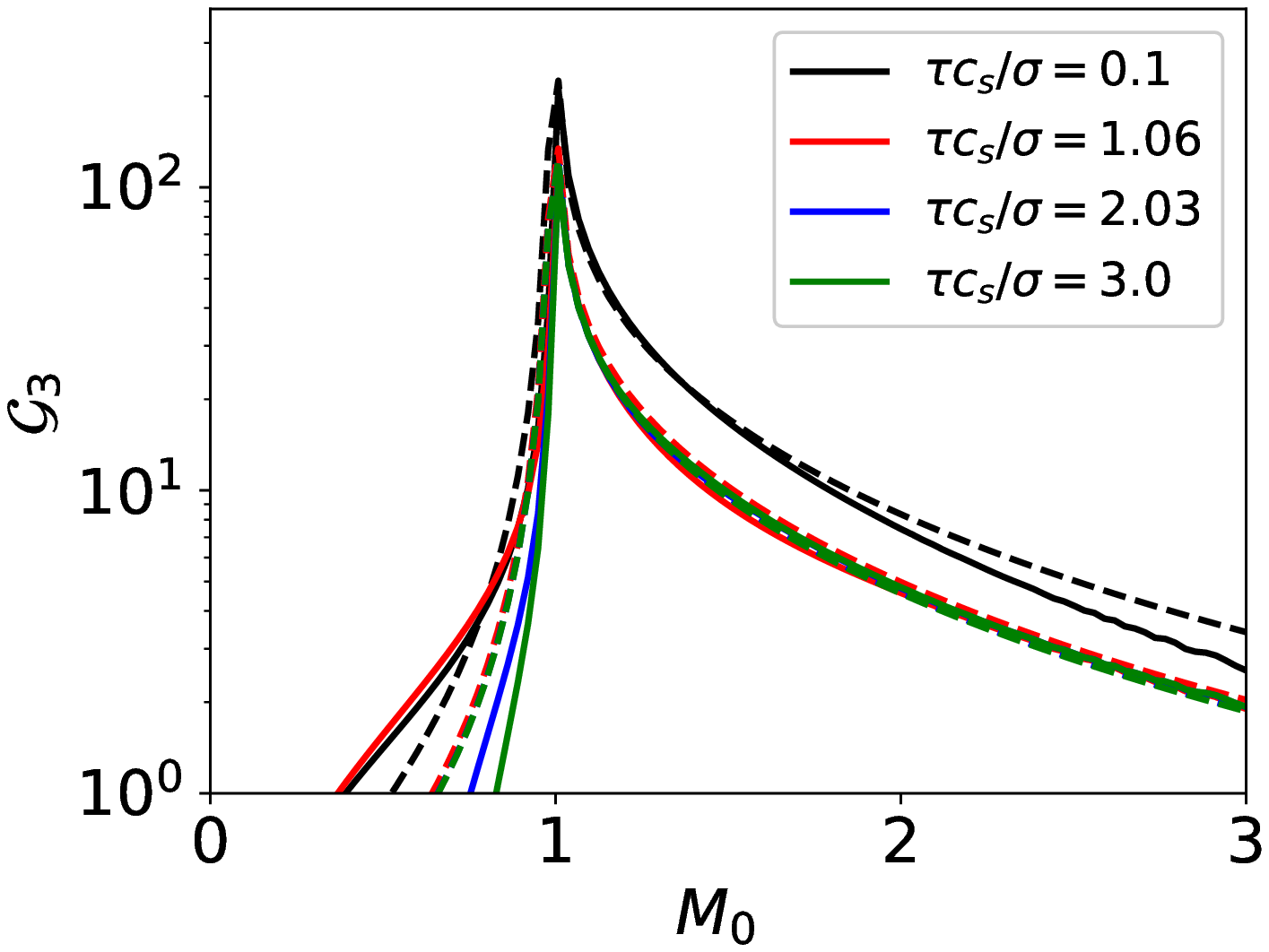}
&\includegraphics[width=0.4\textwidth]{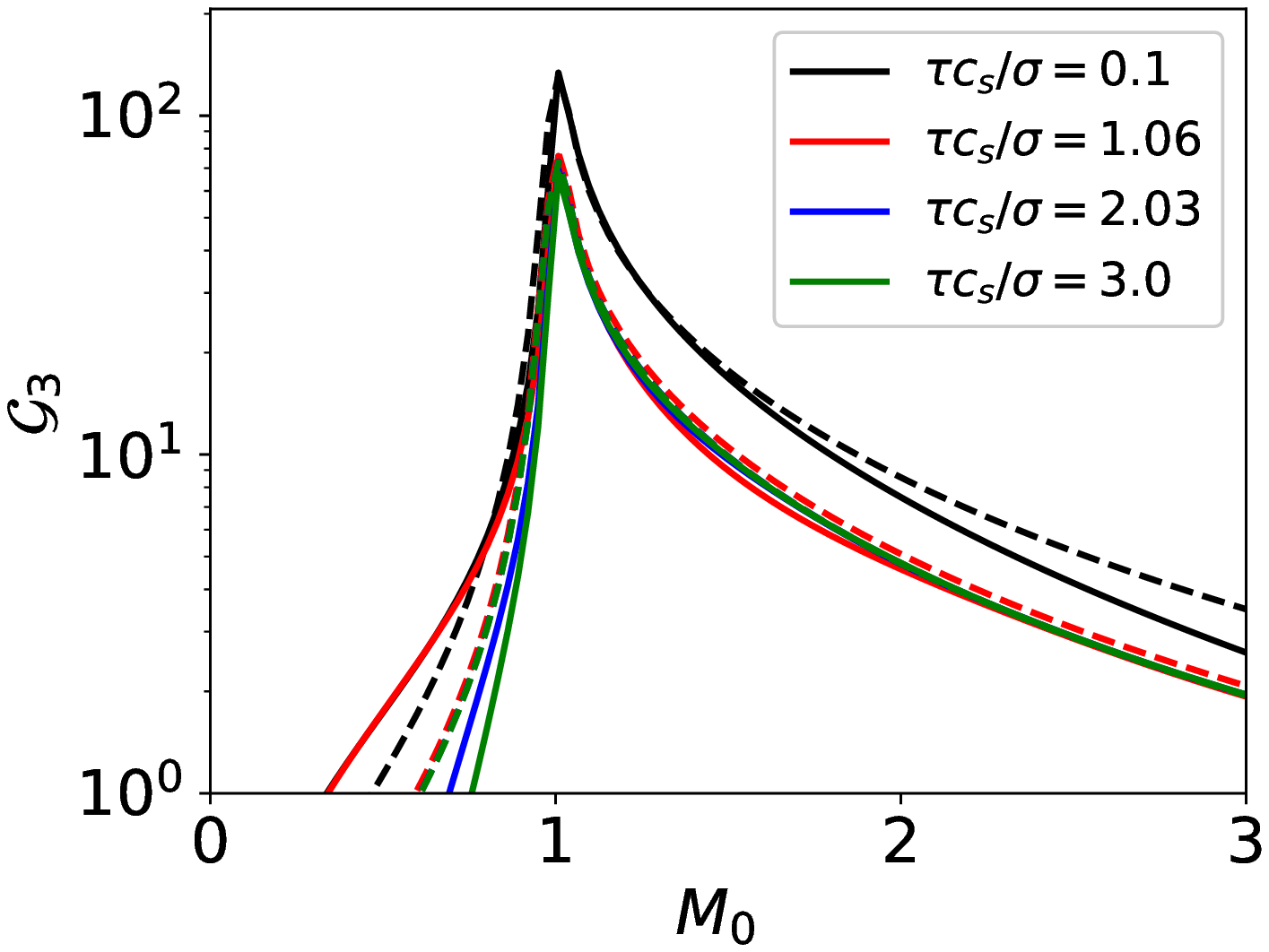}\\
(c)  $\gamma_0=0.05$, Au$^{50+}$ &  (d)  $\gamma_0=0.05$, C$^{6+}$ \\
\includegraphics[width=0.4\textwidth]{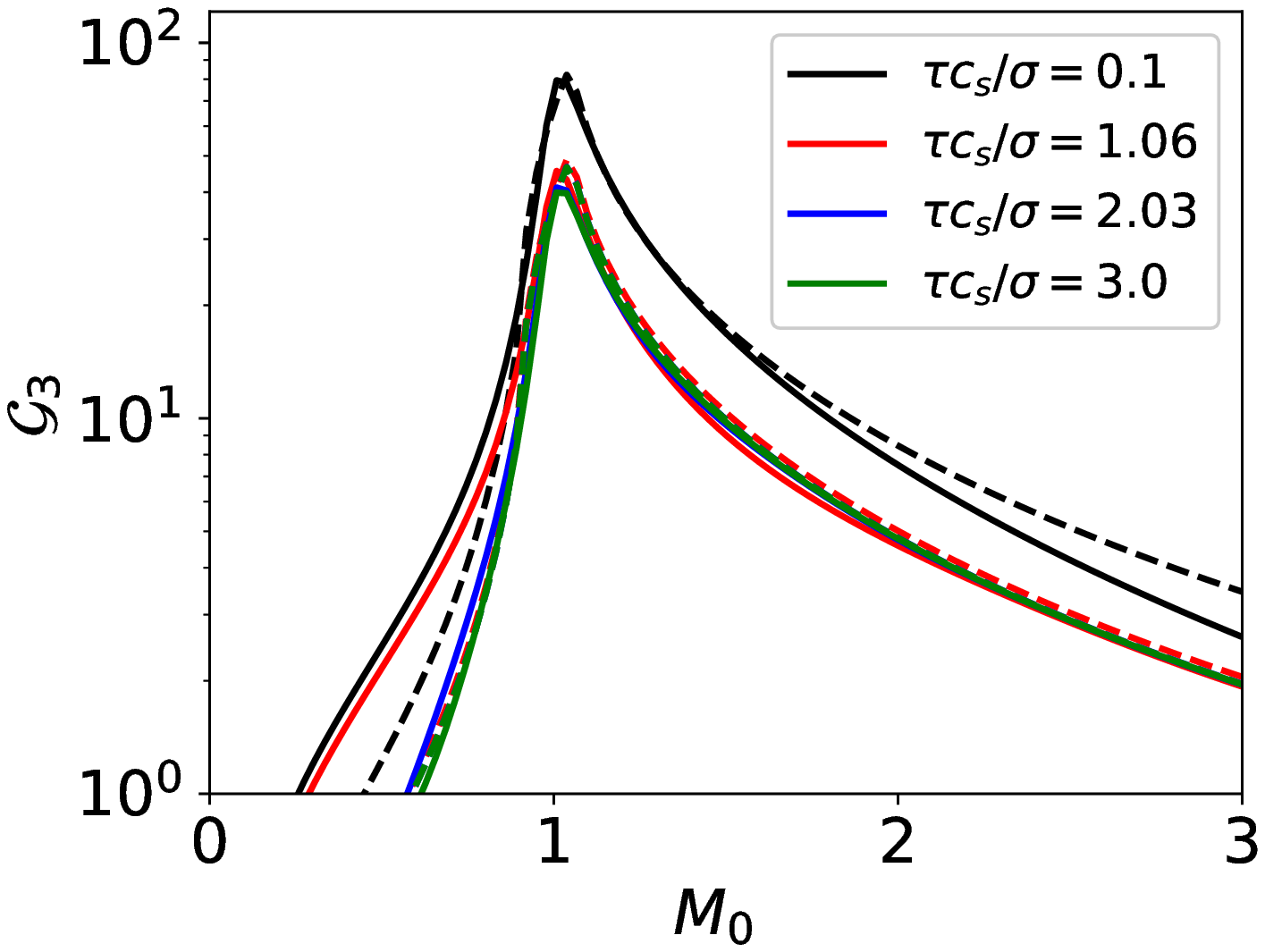}
&\includegraphics[width=0.4\textwidth]{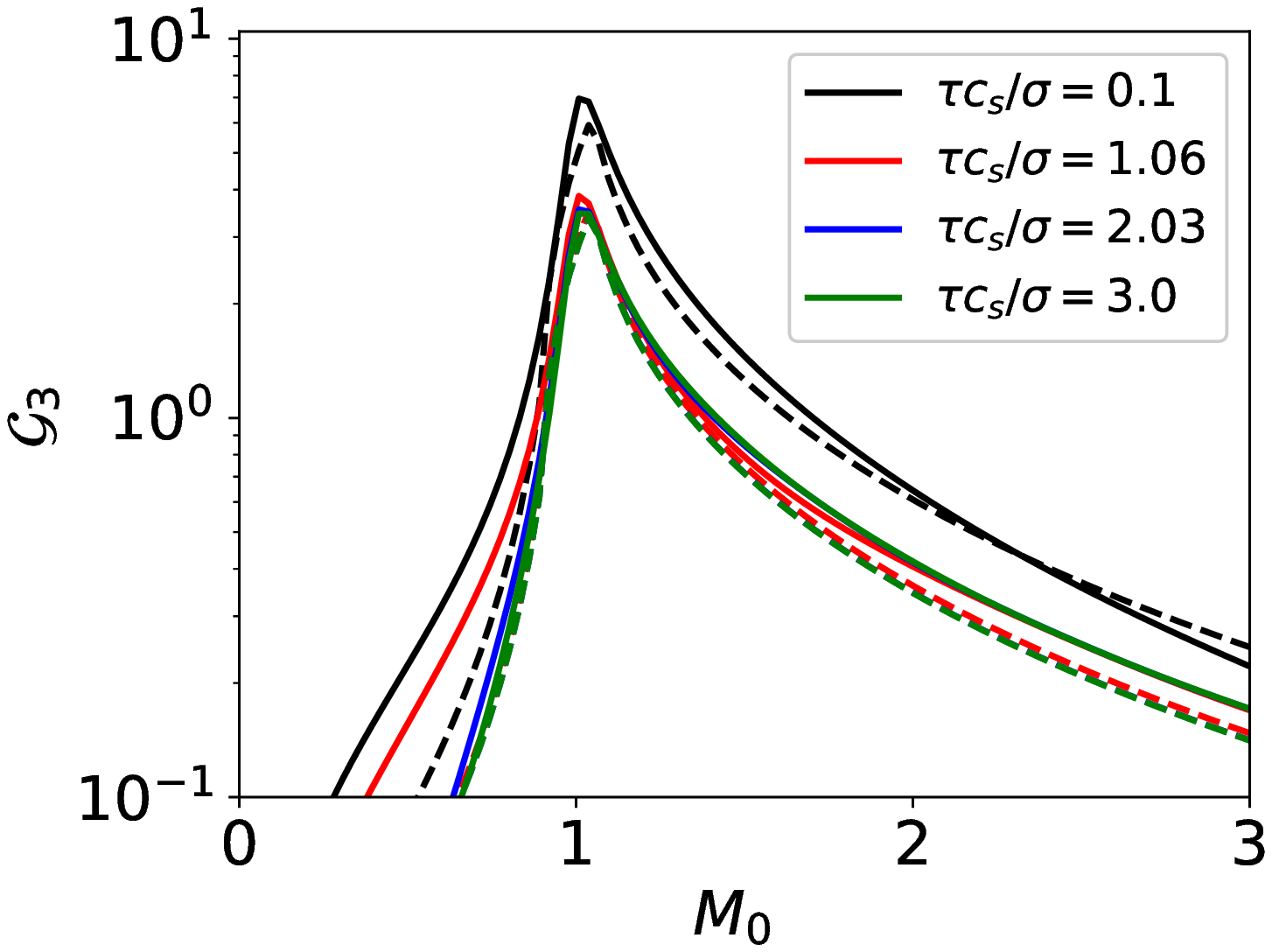}
\end{tabular}
\caption{ \label{fig:fit} 
Comparison between the fit (as dashed lines) and the exact calculations (as plain lines)  for different values of $\tau_\mathrm{SSD}$. The parameters correspond to a gold plasma  (a, b, c, $Z_{Au}=50$, $\lambda_\mathrm{mfp}/f_\sharp\lambda_0 = 1.47$) and carbon (d, $Z_{C}=6$,  $\lambda_\mathrm{mfp}/f_\sharp\lambda_0 = 71)$ with $T_e=3\,\rm keV$, $n_e=9\times 10^{26}\, \rm m^{-3}$.
}
\end{figure*}
In order to facilitate the implementation of the present beam bending model in a hydrodynamic code, we introduce here  a fit of the function $\langle \mathcal{G}  \rangle_\mathrm{\tau_\mathrm{SSD}}$ [Eq. \eqref{eq:G3m}] that appears in the ray tracing model of Eq. \eqref{eq:raybb}. 
Inspired by the analytical model of Ref. \cite[]{POP_Hinkel_1998}, we propose to use the following function,
if $0\le M_0\le 1-\gamma_0$:
\begin{align}
\langle \mathcal{G}  \rangle_\mathrm{\tau_\mathrm{SSD}} &= \frac{A_{k_c}}{8} \frac{M_0}{(1-M_0^2)^{3/2}} F\left(\frac{\tau_\mathrm{SSD}c_s}{f_\sharp\lambda_0},\gamma_0\right)\, .\label{eq:g1}
\end{align}
If $M_0\ge 1+\gamma_0$:
\begin{align}
\langle \mathcal{G}  \rangle_\mathrm{\tau_\mathrm{SSD}} &= A_{k_c} \frac{2M_0}{\sqrt{M_0^2-1}}F\left(\frac{\tau_\mathrm{SSD}c_s}{f_\sharp\lambda_0},\gamma_0\right) u(\gamma_0)\, .\label{eq:g2}
\end{align}
Otherwise:
\begin{align}
\langle \mathcal{G}  \rangle_\mathrm{\tau_\mathrm{SSD}} &= (a+bM_0 +cM_0^2)F\left(\frac{\tau_\mathrm{SSD}c_s}{f_\sharp\lambda_0},\gamma_0\right)\, , \label{eq:g3}
\end{align}
where $k_c=2^{1/2}/(f_\sharp\lambda_0)$.
When $M_0<0$, we propose to use $\langle \mathcal{G}  \rangle_\mathrm{\tau_\mathrm{SSD}}(-M_0)=-\langle \mathcal{G}  \rangle_\mathrm{\tau_\mathrm{SSD}}(M_0)$.
We now introduce $M_\mathrm{min}=1-\gamma_0$, $M_\mathrm{max}=1+\gamma_0$, $\mathcal{G}_\mathrm{min} = \langle \mathcal{G}  \rangle_\mathrm{\tau_\mathrm{SSD}}(M_\mathrm{min}) $, $\mathcal{G}_\mathrm{max} = \langle \mathcal{G}  \rangle_\mathrm{\tau_\mathrm{SSD}}(M_\mathrm{max}) $ and
\begin{align}
    a &= \frac{(\mathcal{G}_\mathrm{min} -\frac{M_\mathrm{min}^2}{2\gamma_0^2} )M_\mathrm{max} -(\mathcal{G}_\mathrm{max} \frac{M_\mathrm{max}^2}{2\gamma_0^2})M_\mathrm{min}  }{M_\mathrm{max}-M_\mathrm{min}}  \, ,\\
    b &= \frac{ \mathcal{G}_\mathrm{min} -\frac{M_\mathrm{min}^2}{2\gamma_0^2} -a }{M_\mathrm{min}} \, ,\\
    c &= \frac{-1}{8\gamma_0^2}\, .
\end{align} 
The functions $F$ and $u$ in Eqs. \eqref{eq:g1},  \eqref{eq:g2} and  \eqref{eq:g3} follows respectively,
\begin{align}
    F\left(\frac{\tau_\mathrm{SSD}c_s}{f_\sharp\lambda_0},\gamma_0\right) &= a_1(\gamma_0) + a_2(\gamma_0)\tanh\left(\frac{\tau_\mathrm{SSD}c_s}{f_\sharp\lambda_0} a_3(\gamma_0)\right)\, , \\
    a_{1,2,3} &= \sum_{n=0}^{2} a_{1,2,3}^{(n)}\gamma_0^n \, . \\
    u &= \sum_{n=0}^{2} u^{(n)}\gamma_0^n \, .
\end{align}
The polynomial coefficients of $a_1(\gamma_0)$, $a_2(\gamma_0)$, $a_3(\gamma_0)$ and  $u(\gamma_0)$ are defined in table \ref{tab:bb}.
\begin{table}[]
    \centering
    \begin{tabular}{|c|c|c|c|}
             \hline
                         & $\times \gamma_0^0$ &  $\times\gamma_0^1$ & $\times\gamma_0^2$ \\
         \hline
         $a_1(\gamma_0)$   &   0.803089 &  30.5832      &  -383.385     \\ 
         \hline
         $a_2(\gamma_0)$   &  -0.397372 &  -14.3873     &  -183.400     \\ 
         \hline
         $a_3(\gamma_0)$   &  1.322688  &  13.1032      &  -100.377 \\ 
         \hline
         $u(\gamma_0)$   &1.749088     &   -32.9549     & 419.593  \\ 
         \hline
    \end{tabular}
    \caption{ \label{tab:bb}
    Polynomial coefficients of  $a_1(\gamma_0)$, $a_2(\gamma_0)$,  $a_3(\gamma_0)$ and  $u(\gamma_0)$.}
\end{table}
This fit works correctly in the domain $0.007 \le \gamma_0\le 0.05$ and  $\vert M_0\vert < 3$ in a mono-species plasma, as shown in Figs. \ref{fig:fit}(a-d).

%\bibliographystyle{aip}
%\bibliography{biblio}
\providecommand{\noopsort}[1]{}\providecommand{\singleletter}[1]{#1}%

\end{document}